\documentclass[epjc3,a4paper,twocolumn]{svjour3}          
\RequirePackage[T1]{fontenc}

\RequirePackage{graphicx}
\RequirePackage{mathptmx}      
\RequirePackage{flushend}
\RequirePackage[numbers,sort&compress]{natbib}
\RequirePackage{amssymb}
\RequirePackage{amsmath}

\journalname{Eur. Phys. J. C}

\usepackage{authblk}
\renewcommand\Authands{, } 
\renewcommand\Affilfont{\itshape\small} 

\usepackage{rotating} 
\usepackage{subfigure}
\usepackage{bm}
\usepackage{atlasphysics}
\usepackage[hyperindex,breaklinks]{hyperref} 

\usepackage{preprintcover}  
\PreprintCoverPaperTitle{Improved luminosity determination in 
$\boldsymbol{pp}$\ collisions at $\mathbf{\sqrt{\boldsymbol{s}} = 7}$~\TeV\ using 
the ATLAS detector at the LHC}  
\PreprintIdNumber{CERN-PH-EP-2013-026}  
\PreprintCoverAbstract{%
The luminosity calibration for the ATLAS detector at the LHC
during $pp$ collisions at $\sqrt s = 7$~\TeV\ in 2010 and 2011 is presented.
Evaluation of the luminosity scale is performed using several luminosity-sensitive 
detectors, and comparisons are made of the long-term stability and accuracy of this 
calibration applied to the $pp$ collisions at $\sqrt s = 7$~\TeV.
A luminosity uncertainty of $\delta {\cal L}/ {\cal L} = \pm 3.5\%$ is obtained for the 
$47\, \mathrm{pb}^{-1}$ of data delivered to ATLAS in 2010, and an uncertainty of  
$\delta {\cal L}/ {\cal L} = \pm 1.8\%$ is obtained for the $5.5\, \mathrm{fb}^{-1}$ 
delivered in 2011.}
\PreprintJournalName{EPJC}  

\begin{document}
\title{Improved luminosity determination in $\boldsymbol{pp}$\ collisions at $\mathbf{\sqrt{\boldsymbol{s}} = 7}$~\TeV\ using the ATLAS detector at the LHC}
\author{\bf The ATLAS Collaboration}
\institute{CERN, 1211 Geneve 23, Switzerland}
\date{Received: date / Accepted: date}
%
\maketitle
%
\begin{abstract}
The luminosity calibration for the ATLAS detector at the LHC
during $pp$ collisions at $\sqrt s = 7$~\TeV\ in 2010 and 2011 is presented.
Evaluation of the luminosity scale is performed using several luminosity-sensitive 
detectors, and comparisons are made of the long-term stability and accuracy of this 
calibration applied to the $pp$ collisions at $\sqrt s = 7$~\TeV.
A luminosity uncertainty of $\delta {\cal L}/ {\cal L} = \pm 3.5\%$ is obtained for the 
$47\, \mathrm{pb}^{-1}$ of data delivered to ATLAS in 2010, and an uncertainty of  
$\delta {\cal L}/ {\cal L} = \pm 1.8\%$ is obtained for the $5.5\, \mathrm{fb}^{-1}$ 
delivered in 2011.
\end{abstract}

\section{Introduction}
An accurate measurement of the delivered luminosity is a key component of the 
ATLAS~\cite{bib:ATLASDetectorPaper} physics programme.
For cross-section measurements, the uncertainty on the delivered luminosity is often one 
of the major systematic uncertainties.
Searches for, and eventual discoveries of, new physical phenomena beyond the 
Standard Model also rely on accurate information about the delivered luminosity to 
evaluate background levels and determine sensitivity to the signatures of new 
phenomena.

\begin{figure}[htbp] 
   \includegraphics[width=\columnwidth]{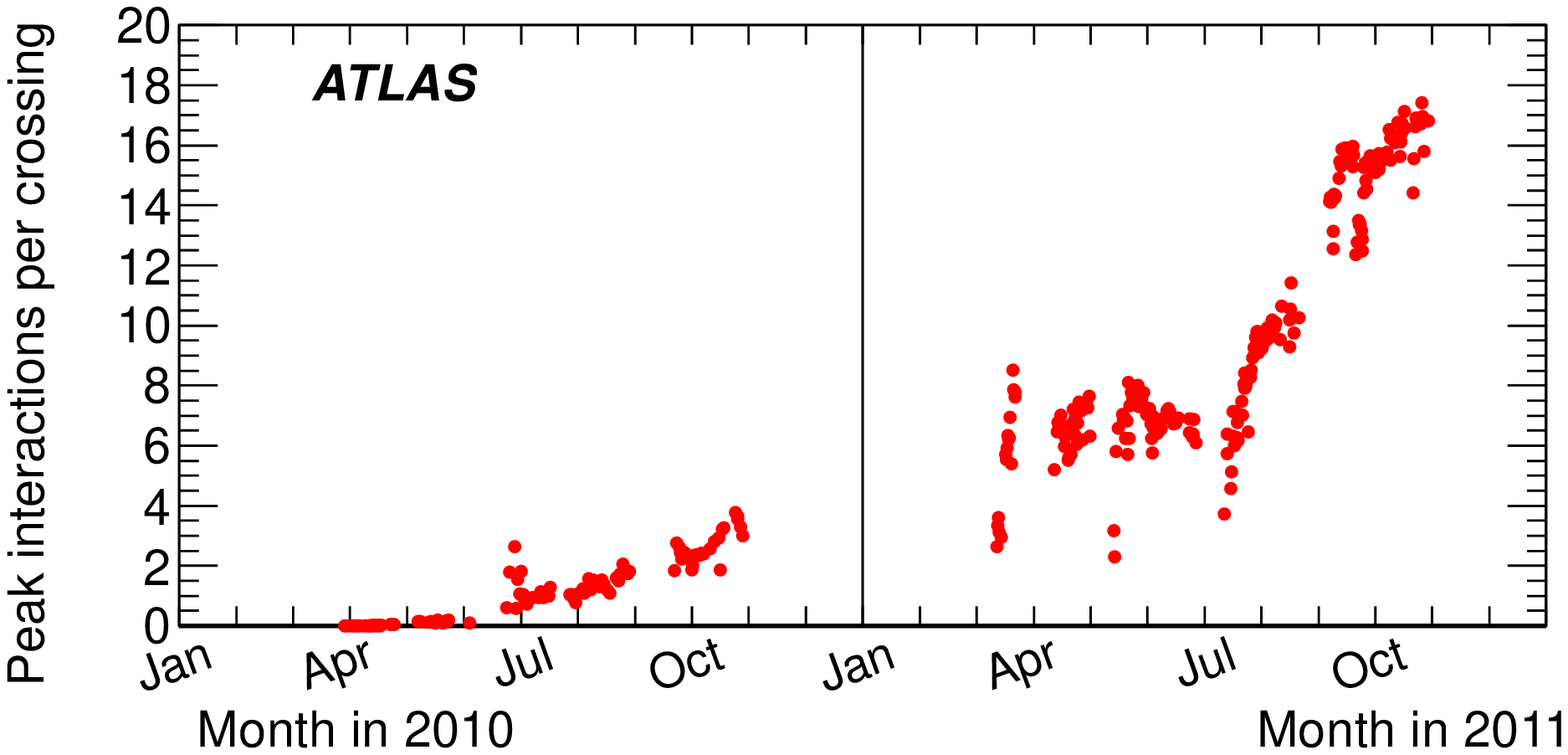} 
   \includegraphics[width=\columnwidth]{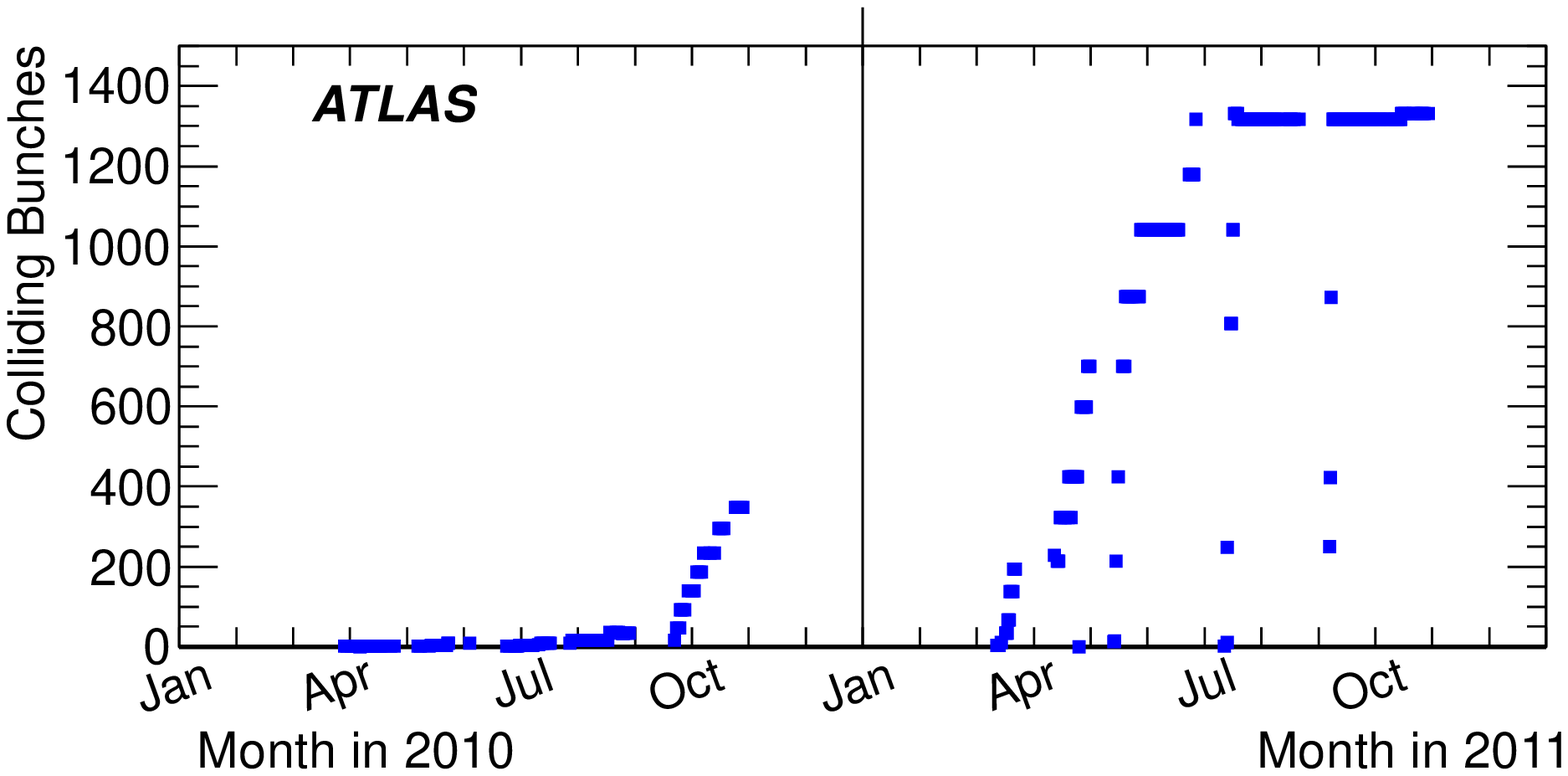} 
   \caption{Average number of inelastic $pp$ interactions per bunch crossing at the start of each LHC fill (above) and number of colliding bunches per LHC fill (below) are shown as a function of time in 2010 and 2011.   The product of these two quantities is proportional to the peak luminosity at the start of each fill.}
   \label{fig:interactions}
\end{figure}

This paper describes the measurement of the luminosity delivered to the ATLAS detector 
at the LHC in $pp$ collisions at a centre-of-mass energy of $\sqrt{s}=7$~\TeV\ during 
2010 and 2011.
The analysis is an evolution of the process documented in the initial ATLAS luminosity 
publication~\cite{bib:ATLASLumiPaper} and includes an improved determination of 
the luminosity in 2010 along with a new analysis for 2011.
Table~\ref{tab:LHC} highlights the operational conditions of the LHC during 2010 
and 2011.  
The peak instantaneous luminosity delivered by the LHC at the start of a fill increased 
from ${\cal L}_{\mathrm{peak}} =  2.0 \times 10^{32}\, \mathrm{cm}^{-2}\, \mathrm{s}^{-1}$ 
in 2010 to 
${\cal L}_{\mathrm{peak}} =  3.6 \times 10^{33}\, \mathrm{cm}^{-2}\, \mathrm{s}^{-1}$ 
by the end of 2011.
This increase results from both an increased instantaneous luminosity delivered per 
bunch crossing as well as a significant increase in the total number of bunches colliding.
Figure~\ref{fig:interactions} illustrates the evolution of these two parameters as a function 
of time.
As a result of these changes in operating conditions, the details of the luminosity 
measurement have evolved from 2010 to 2011, although the overall methodology 
remains largely the same.

\begin{table}
\centering
\caption{Selected LHC parameters for $pp$ collisions at $\sqrt{s} = 7$~\TeV\ in 2010 and 2011. Parameters shown are the best achieved for that year in normal physics operations.}
\label{tab:LHC}
\begin{tabular*}{\columnwidth}{@{\extracolsep{\fill}}lcc@{}}
      	\hline
      	Parameter & 2010 & 2011 \\
      	\hline
	Maximum number of bunch pairs colliding & 348 & 1331 \\
	Minimum bunch spacing (ns) & 150 & 50 \\
	Typical bunch population ($10^{11}$ protons) & $0.9$ & $1.2$ \\
	Peak luminosity $(10^{33}\, \mathrm{cm}^{-2}\, \mathrm{s}^{-1})$ & 0.2 & 3.6 \\
	Maximum inelastic interactions per crossing & $\sim 5$ & $\sim 20$ \\
	Total integrated luminosity delivered & $47\, \mathrm{pb}^{-1}$ & $5.5\, \mathrm{fb}^{-1}$ \\ 
	\hline
   \end{tabular*}
\end{table}

The strategy for measuring and calibrating the luminosity is outlined  
in Sect.~\ref{sec:overview}, followed in Sect.~\ref{sec:detectors} by a brief description of 
the detectors used for luminosity determination.
Each of these detectors utilizes one or more luminosity algorithms as
described in Sect.~\ref{sec:algorithms}. 
The absolute calibration of these algorithms using
beam-separation scans is described in Sect.~\ref{sec:calibration}, 
while a summary of  the systematic uncertainties on the luminosity calibration 
as well as the calibration results are presented in Sect.~\ref{sec:errors}.  
Additional corrections which must be applied over the course of the 2011 data-taking 
period are described in Sect.~\ref{sec:extrapolation}, while 
additional uncertainties related to the extrapolation of the absolute luminosity calibration 
to the full 2010 and 2011 data samples are described in Sect.~\ref{sec:stability}.
The final results and uncertainties are summarized in Sect.~\ref{sec:conclusions}.

\section{Overview}
\label{sec:overview}

The  luminosity ${\cal L}$ of a $pp$ collider can be expressed as
\begin{equation}
\mathcal{L} = \frac{R_{\mathrm{inel}}}{\sigma_{\mathrm{inel}}}
\end{equation}
where  $R_{\mathrm{inel}}$ is the rate of inelastic collisions and $\sigma_{\mathrm{inel}}$ is
the $pp$ inelastic cross-section.  
For a storage ring, operating at a revolution frequency $f_\mathrm{r}$ and with $n_\mathrm{b}$ 
bunch pairs colliding per revolution,  this expression can be rewritten as
\begin{equation}
\mathcal{L} = \frac{{\mu n_\mathrm{b} f_\mathrm{r} }}{{\sigma _{\mathrm{inel}} }}
\label{eq:lumiToMu}
\end{equation}
where $\mu$ is the average number of inelastic interactions per 
bunch crossing. 

As discussed in Sects.~\ref{sec:detectors} and \ref{sec:algorithms}, ATLAS monitors 
the delivered luminosity by measuring the observed interaction rate per crossing, 
$\mu_{\mathrm{vis}}$, independently with a variety of detectors and using several 
different algorithms.
The luminosity can then be written as
\begin{equation}
\mathcal{L} = \frac{{\mu_{\mathrm{vis}} n_{\mathrm{b}} f_{\mathrm{r}} }}{{\sigma _{\mathrm{vis}} }}
\label{eqn:defmu}
\end{equation}
where $\sigma _{\mathrm{vis}} = \varepsilon \sigma _{\mathrm{inel}}$ is the total inelastic cross-section multiplied by the efficiency $\varepsilon$ of a particular detector and algorithm, and similarly $\mu_{\mathrm{vis}} = \varepsilon \mu$.
Since $\mu_{\mathrm{vis}}$ is an experimentally observable quantity, 
the calibration of the luminosity scale for a particular detector and algorithm
is equivalent to determining the visible cross-section $\sigma_{\mathrm{vis}}$.

The majority of the algorithms used in the ATLAS luminosity determination are {\em event counting} 
algorithms, where each particular bunch crossing is categorized as either passing or not passing
a given set of criteria designed to detect the presence of at least one inelastic $pp$ collision.
In the limit $\mu_{\mathrm{vis}} \muchless 1$, the average number of visible inelastic 
interactions per bunch crossing is given by the simple expression
$\mu_{\mathrm{vis}} \approx N / N_{\mathrm{BC}}$
where $N$ is the number of bunch crossings (or events) passing the selection criteria that are 
observed during a given time interval, and $N_\mathrm{BC}$ is the total number of bunch crossings 
in that same interval.
As $\mu_{\mathrm{vis}}$ increases, the probability that two 
or more $pp$ interactions occur in the same bunch crossing is no longer 
negligible (a condition referred to as ``pile-up''), 
and $\mu_{\mathrm{vis}}$ is no longer linearly related to the raw event count $N$. 
Instead  $\mu_{\mathrm{vis}}$ must be calculated taking into account Poisson statistics, 
and in some cases instrumental or pile-up-related effects.
In the limit where all bunch crossings in a given time interval contain an event, the event counting
algorithm no longer provides any useful information about the interaction rate.

An alternative approach, which is linear to higher values of $\mu_{\mathrm{vis}}$ but requires control of additional
systematic effects, is that of {\em hit counting} algorithms.
Rather than counting how many bunch crossings pass some minimum criteria for  containing at least one inelastic interaction, 
in hit counting algorithms the number
of detector readout channels with signals above some predefined threshold is counted. 
This provides more information per event, and also increases the $\mu_{\mathrm{vis}}$ value at which the algorithm saturates compared 
to an event-counting algorithm.
The extreme limit of hit counting algorithms, achievable only in detectors with very fine segmentation, 
are {\em particle counting} algorithms, where the number of individual particles entering a given detector is counted directly.
More details on how these different algorithms are defined, as well as the procedures for converting the observed event or hit rate into the
visible interaction rate $\mu_{\mathrm{vis}}$, are discussed in Sect.~\ref{sec:algorithms}.

As described more fully in Sect.~\ref{sec:calibration}, the calibration 
of $\sigma_{\mathrm{vis}}$ is performed using dedicated beam-separation scans, also known
as van der Meer ({\it vdM}) scans, where the absolute luminosity can be inferred from direct 
measurements of the beam parameters~\cite{bib:vdm, bib:Rubbia}.
The delivered luminosity can be written in terms of the accelerator parameters as
\begin{equation}
{\mathcal L} = \frac{{n_{\mathrm{b}} f_{\mathrm{r}} n_1 n_2 }}{{2\pi \Sigma _x \Sigma _y
}}\label{eqn:defLumi}
\end{equation}
where $n_1$ and $n_2$ are the bunch populations (protons per bunch) in beam 1 and beam 2 
respectively (together forming the bunch population product), 
and $\Sigma_x$ and $\Sigma_y$ characterize 
the horizontal and vertical convolved beam widths. 
In a {\it vdM} scan, the beams are separated by steps of a known distance, which allows a 
direct measurement of $\Sigma _x$ and $\Sigma _y$.
Combining this scan with an external measurement of the bunch population product 
$n_1 n_2$ provides a direct determination of the luminosity when the beams are unseparated.

A fundamental ingredient of the ATLAS strategy to assess and control the 
systematic uncertainties affecting the absolute luminosity determination is to 
compare the measurements of several luminosity detectors, most of which use 
more than one algorithm to assess the luminosity. 
These multiple detectors and algorithms are characterized by 
significantly different acceptance, response to pile-up, 
and sensitivity to instrumental effects and to beam-induced backgrounds. 
In particular, since the calibration of the absolute luminosity scale is established in dedicated
{\em vdM} scans which are carried out relatively infrequently (in 2011 there was only one set of {\em vdM}
scans at $\sqrt s = 7$~\TeV\ for the entire year), this calibration must be assumed to be constant over long  
periods and under different machine conditions.
The level of consistency across the various methods, over the full range of 
single-bunch luminosities and beam conditions, and across many months of LHC operation,
provides valuable cross-checks as well as an estimate of the detector-related 
systematic uncertainties. A full discussion of these is
presented in Sects.~\ref{sec:errors}--\ref{sec:stability}.

The information needed for most physics analyses is an integrated luminosity for some well-defined data sample.
The basic time unit for storing luminosity information for physics use is the Luminosity Block (LB). 
The boundaries of each LB are defined by the ATLAS Central Trigger Processor (CTP), and in general the duration of each LB is 
one minute.
Trigger configuration changes, such as prescale changes, can only happen at luminosity block boundaries, and data are 
analysed under the assumption that 
each luminosity block contains data taken under uniform conditions, including luminosity.
The average luminosity for each detector and algorithm, along with a variety of general ATLAS data quality information, is stored for 
each LB in a relational database.
To define a data sample for physics, quality criteria are applied to select LBs where conditions are acceptable, then the 
average luminosity in that LB is multiplied by the LB duration to provide the integrated luminosity delivered in that LB.
Additional corrections can be made for trigger deadtime and trigger prescale factors, which are also recorded on a per-LB basis.
Adding up the integrated luminosity delivered in a specific set of luminosity blocks provides the integrated luminosity of the entire data sample.

\section{Luminosity detectors}
\label{sec:detectors}

This section provides a description of the detector subsystems used for 
luminosity measurements.
The ATLAS detector is discussed in detail in Ref.~\cite{bib:ATLASDetectorPaper}.  
The first set of detectors uses either event or hit counting 
algorithms to measure the luminosity on a bunch-by-bunch basis.  
The second set infers the total luminosity (summed over all bunches) by monitoring 
detector currents sensitive to average particle rates over longer time scales.
In each case, the detector descriptions are arranged in order of increasing magnitude 
of pseudorapidity.\footnote{
ATLAS uses a right-handed coordinate system with its origin at the nominal interaction point (IP) in the centre of the detector, and the $z$-axis along the beam line.  The $x$-axis points from the IP to the centre of the LHC ring, and the $y$-axis points upwards.
Cylindrical coordinates $(r, \phi)$ are used in the transverse plane, $\phi$ being the azimuthal angle around the beam line.  The pseudorapidity is defined in terms of the polar angle $\theta$ as $\eta = -\ln \tan(\theta/2)$.}

The Inner Detector is used to measure the momentum of charged particles over a
pseudorapidity interval of $|\eta| < 2.5$.
It consists of three subsystems: a pixel detector, a silicon microstrip tracker, and a 
transition-radiation straw-tube tracker.  
These detectors are located inside a solenoidal magnet that provides a 2~T axial field.  
The tracking efficiency as a function of transverse momentum ($p_\mathrm{T}$), averaged over
all pseudorapidity, rises from $10\%$ at 100 MeV to around $86\%$
for $p_\mathrm{T}$ above a few \GeV~\cite{Aad:2010ac, bib:CONF2012-042}.
The main application of the Inner Detector for luminosity measurements is to detect
the primary  vertices produced in inelastic $pp$ interactions.

To provide efficient triggers at low instantaneous luminosity (${\cal L} < 10^{33}~{\rm
cm}^{-2}{\rm s}^{-1}$), ATLAS has been equipped with segmented
scintillator counters, the Minimum Bias Trigger Scintillators (MBTS).
Located at $z=\pm365$~cm from the nominal interaction point (IP),
and covering a rapidity range $2.09 < |\eta| < 3.84$,
the main purpose of the MBTS system is to provide a trigger on minimum collision activity 
during a $pp$  bunch crossing.
Light emitted by the scintillators is collected by wavelength-shifting optical fibers and 
guided to photomultiplier tubes.
The MBTS signals, after being shaped and amplified, are fed into leading-edge
discriminators and sent to the trigger system.  
The MBTS detectors are primarily used for luminosity measurements in early 2010, and
are no longer used in the 2011 data.

The Beam Conditions Monitor (BCM) consists of four small  diamond sensors, approximately 
1 cm$^2$ in cross-section each, arranged around the beampipe in a cross pattern on each side of the 
IP, at a distance of $z = \pm 184$~cm.
The BCM is a fast device originally designed to monitor background levels and issue 
beam-abort requests when beam losses start to risk damaging the Inner Detector.
The fast readout of the BCM also provides a bunch-by-bunch 
luminosity signal at $|\eta| = 4.2$ with a time resolution of $\simeq 0.7$~ns.
The horizontal and vertical pairs of BCM detectors are read out separately, leading to two  
luminosity measurements labelled BCMH and BCMV respectively.
Because the acceptances, thresholds, and data paths may all have small differences between 
BCMH and BCMV, these two measurements are treated as being made by independent devices for 
calibration and monitoring purposes, although the overall response of the two devices 
is expected to be very similar.
In the 2010 data, only the BCMH readout is available for luminosity measurements, while 
both BCMH and BCMV are available in 2011.

LUCID is a Cherenkov detector specifically designed for measuring the 
luminosity. 
Sixteen mechanically polished aluminium tubes filled 
with ${\rm C}_4{\rm F}_{10}$ gas surround 
the beampipe on each side of the IP at a distance of 17~m,
covering the pseudorapidity range $5.6 < |\eta| < 6.0$.
The Cherenkov photons created by charged particles in the gas are reflected by the 
tube walls until they reach photomultiplier tubes (PMTs) situated at the back end of the tubes.
Additional Cherenkov photons are produced in the quartz window separating the 
aluminium tubes from the PMTs.
The Cherenkov light created in the gas typically produces 60--70 photoelectrons per incident
charged particle, while the quartz window adds another 40 photoelectrons to the signal.
If one of the LUCID PMTs produces a signal over a preset threshold (equivalent to 
$\simeq 15$ photoelectrons), a ``hit'' is recorded for that tube in that bunch crossing.
The LUCID hit pattern is processed by a custom-built electronics card which
contains Field Programmable Gate Arrays (FPGAs). 
This card can be programmed with 
different luminosity algorithms, and provides separate luminosity measurements for 
each LHC bunch crossing.

Both BCM and LUCID are fast detectors with electronics capable of making statistically 
precise luminosity measurements separately for each bunch crossing within the LHC fill 
pattern with no deadtime.
These FPGA-based front-end electronics run autonomously from the main data 
acquisition system, and in particular are not affected by any deadtime imposed by 
the CTP.\footnote{The CTP inhibits triggers (causing deadtime) for a variety of reasons, 
but especially for several bunch crossings after a triggered event to allow time for the 
detector readout to conclude.  
Any new triggers which occur during this time are ignored.}

The Inner Detector vertex data and the MBTS data are components of the events read out through the  
data acquisition system, and so must be corrected for deadtime imposed by the CTP
in order to measure delivered luminosity.  Normally this deadtime is below 1\%, but can 
occasionally be larger.
Since not every inelastic collision event can be read out through the data acquisition 
system, 
the bunch crossings are sampled with a random or minimum bias trigger.
While the triggered events uniformly sample every bunch crossing, the trigger 
bandwidth devoted to random or minimum bias triggers is not large enough to measure the 
luminosity separately for each bunch pair in a given LHC fill pattern during normal physics 
operations.
For special running conditions such as the {\em vdM} scans, a custom trigger with partial
event readout has been introduced in 2011 to record enough events to allow bunch-by-bunch luminosity 
measurements from the Inner Detector vertex data.

In addition to the detectors listed above, further luminosity-sensitive methods have 
been developed which use components of the ATLAS calorimeter system.  
These techniques do not identify particular events, but rather measure average particle 
rates over longer time scales.  

The Tile Calorimeter (TileCal) is the central hadronic calorimeter of ATLAS. 
It is a sampling calorimeter constructed from iron plates (absorber) and plastic tile scintillators 
(active material) covering the pseudorapidity range $|\eta| < 1.7$.
The detector consists of three cylinders, a central long barrel and two smaller extended
barrels, one on each side of the long barrel. 
Each cylinder is divided into 64 slices in $\phi$ (modules) and segmented into three radial sampling layers.
Cells are defined in each layer according to a projective geometry, and each
cell is connected by optical fibers to two photomultiplier tubes.
The current drawn by each PMT is monitored by an integrator system which is sensitive to 
currents from 0.1~nA to 1.2~mA with a time constant of 10~ms.
The current drawn is proportional to the total number of particles interacting in a given
TileCal cell, and provides a signal proportional to the total luminosity summed over all  
the colliding bunches present at a given time.
 
The Forward Calorimeter (FCal) is a sampling calorimeter that covers the pseudorapidity 
range $3.2 < |\eta| < 4.9$ and is housed in the two endcap cryostats along with the 
electromagnetic endcap and the hadronic endcap calorimeters. 
Each of the two FCal modules is divided into three longitudinal absorber matrices, one made of 
copper (FCal-1) and the other two of tungsten (FCal-2/3). 
Each matrix contains tubes arranged parallel to the beam axis filled with liquid argon as the active 
medium. 
Each FCal-1 matrix is divided into 16 $\rm\phi$-sectors, each of them fed by four independent 
high-voltage lines.
The high voltage on each sector is regulated to provide a stable electric field across the liquid 
argon gaps and, similar to the TileCal PMT currents, the currents provided by the FCal-1 
high-voltage system are directly proportional to the average rate of particles interacting in a given 
FCal sector.

\section{Luminosity algorithms}
\label{sec:algorithms}
This section describes the algorithms used by the luminosity-sensitive detectors described in Sect.~\ref{sec:detectors} to measure the visible interaction rate per bunch crossing, 
$\mu_\mathrm{vis}$.
Most of the algorithms used do not measure $\mu_\mathrm{vis}$ directly, but rather measure 
some other rate which can be used to determine $\mu_\mathrm{vis}$.

ATLAS primarily uses event counting algorithms to measure luminosity, 
where a bunch crossing is said to contain an ``event'' if the criteria for a given 
algorithm to observe one or more interactions are satisfied.
The two main algorithm types being used are 
EventOR (inclusive counting) and EventAND (coincidence counting).
Additional algorithms have been developed using hit counting and average particle rate 
counting, which provide a cross-check of the linearity of the event counting techniques.

\subsection{Interaction rate determination}
\label{sec:mu}

Most of the primary luminosity detectors consist of two symmetric detector elements 
placed in the forward (``A'') and backward (``C'') direction from the interaction point.
For the LUCID, BCM, and MBTS detectors, each side is further segmented into a discrete number
of readout segments, typically arranged azimuthally around the beampipe, each with a separate
readout channel.
For event counting algorithms, a threshold is applied to the analoge signal output from each readout
channel, and every channel with a response above this threshold is counted as containing a ``hit''.

In an EventOR algorithm, a bunch crossing is counted if there is at least one hit on 
either the A side or the C side.
Assuming that the number of interactions in a bunch crossing can be described by a Poisson 
distribution, the probability of observing an OR event can be 
computed as
\begin{equation}
P_\textsf{\tiny Event\_OR}(\mu_{\mathrm{vis}}^{\mathrm{OR}})  =  \frac{N_{\mathrm{OR}}}{N_{\mathrm{BC}}} = 1-e^{-\mu_{\mathrm{vis}}^{\mathrm{OR}}}.
\label{eqn:muor}
\end{equation}
Here the raw event count $N_{\mathrm{OR}}$ is the number of bunch crossings, 
during a given time interval, in which at least one $pp$ interaction satisfies the event-selection 
criteria of the OR algorithm under consideration, and $N_{\mathrm{BC}}$ is the total 
number of bunch crossings during the same interval.
Solving for $\mu_{\mathrm{vis}}$ in terms of the event counting rate yields:
\begin{equation}
\mu_{\mathrm{vis}}^{\mathrm{OR}} = - \ln \left( 1- \frac{N_{\mathrm{OR}}}{N_{\mathrm{BC}}} \right ).\\
\label{eqn:muand}
\end{equation}

In the case of an EventAND algorithm, a bunch crossing is counted if there is at least one hit 
on both sides of the detector.
This coincidence condition can be satisfied either from a single $pp$ interaction or from individual 
hits on either side of the detector from different $pp$ interactions in the same bunch crossing.
Assuming equal acceptance for sides A and C, the probability of recording an AND event can be expressed as
\begin{equation}
\begin{array}{lclcl}
P_\textsf{\tiny Event\_AND}(\mu_{\mathrm{vis}}^{\mathrm{AND}}) & = & \frac{N_{\mathrm{AND}}}{N_{\mathrm{BC}}}\\
            & = & 1- 2e^{-(1+ {\sigma_{\mathrm{vis}}^{\mathrm{OR}}}/{\sigma_{\mathrm{vis}}^{\mathrm{AND}}})\mu_{\mathrm{vis}}^{\mathrm{AND}}/2}\\
           & & + e^{-({\sigma_{\mathrm{vis}}^{\mathrm{OR}}}/{\sigma_{\mathrm{vis}}^{\mathrm{AND}}})\mu_{\mathrm{vis}}^{\mathrm{AND}}}.
\end{array}
\label{eq:eventAndProb1}
\end{equation}
This relationship cannot be inverted analytically to determine  $\mu_{\mathrm{vis}}^{\mathrm{AND}}$ as a function of $N_{\mathrm{AND}}/N_{\mathrm{BC}}$ so a numerical inversion is performed instead.

When $\mu_{\mathrm{vis}} \gg 1$, event counting algorithms lose sensitivity as fewer and 
fewer events in a given time interval have bunch crossings with zero observed interactions.
In the limit where $N/N_{\mathrm{BC}} = 1$, it is no longer possible to use event counting to
determine the interaction rate $\mu_{\mathrm{vis}}$, and more sophisticated techniques must
be used.
One example is a {\em hit counting} algorithm, where the number of hits in a given detector is counted rather than just the total number of events.  
This provides more information about the interaction rate per event, and increases 
the luminosity at which the algorithm saturates.

Under the assumption that the number of hits in one $pp$ interaction 
follows a Binomial distribution and that the number of interactions per bunch crossing
follows a Poisson distribution, one can calculate the average probability
to have a hit in one of the detector channels per bunch crossing as
\begin{equation}
P_\textsf{\tiny HIT}(\mu_\mathrm{vis}^{\mathrm{HIT}}) = \frac{N_{\mathrm{HIT}}}{N_\mathrm{BC} N_\mathrm{CH}}
= 1 - e^{-\mu_\mathrm{vis}^{\mathrm{HIT}}},
\end{equation}
where $N_\mathrm{HIT}$ and $N_\mathrm{BC}$ are the total numbers of hits and bunch crossings during a time interval, 
and $N_\mathrm{CH}$ is the number of detector channels. 
The expression above enables $\mu_\mathrm{vis}^{\mathrm{HIT}}$ to be calculated
from the number of hits as
\begin{equation}
\begin{array}{ll}
\mu_\mathrm{vis}^{\mathrm{HIT}}  =  -\ln(1-\frac{N_\mathrm{HIT}}{N_\mathrm{BC}N_\mathrm{CH}}).\\  
\end{array}
\label{eqn:hitor} 
\end{equation}

Hit counting is  used to analyse the LUCID response ($N_\mathrm{CH} = 30$) only in the high-luminosity data taken in 2011.
The lower acceptance of the BCM detector allows event counting to remain viable for all of 2011.
The binomial assumption used to derive Eq.~(\ref{eqn:hitor}) is only true if the probability 
to observe a hit in a single channel is independent of the number of hits observed in the 
other channels. 
A study of the LUCID hit distributions shows that this is not a correct assumption, 
although the data presented in Sect.~\ref{sec:stability} also show that 
Eq.~(\ref{eqn:hitor}) provides a good description of how 
$\mu_\mathrm{vis}^{\mathrm{HIT}}$ depends on the average number of hits.

An additional type of algorithm that can be used is a {\em particle counting} algorithm, where some 
observable is directly proportional to the number of particles interacting in the detector.
These should be the most linear of all of the algorithm types, and in principle the interaction 
rate is directly proportional to the particle rate.  
As discussed below, the TileCal and FCal current measurements are not exactly particle counting
algorithms, as individual particles are not counted, but the measured currents should be directly
proportional to luminosity.  
Similarly, the number of primary vertices is directly proportional to the luminosity, although the vertex
reconstruction efficiency is significantly affected by pile-up as discussed below.

\subsection{Online algorithms}
The two main luminosity detectors used are LUCID and BCM.
Each of these is equipped with customized FPGA-based readout electronics which allow 
the luminosity algorithms to be applied ``online'' in real time.
These electronics provide fast diagnostic signals to the LHC (within a few seconds),
in addition to providing luminosity measurements for physics use.
Each colliding bunch pair can be identified numerically by a Bunch-Crossing Identifier  
(BCID) which labels each of the 3564 possible 25~ns slots in one full revolution of the 
nominal LHC fill pattern.
The online algorithms measure the delivered luminosity independently in each BCID.

For the LUCID detector, the two main algorithms are the inclusive 
LUCID\_EventOR and the coincidence \mbox{LUCID\_EventAND}.
In each case, a hit is defined as a PMT signal above a predefined threshold which is set lower
than the average single-particle response.
There are two additional algorithms defined, LUCID\_EventA and LUCID\_EventC, which require
at least one hit on either the A or C side respectively.
Events passing these LUCID\_EventA and LUCID\_EventC algorithms are  subsets of the 
events passing the LUCID\_EventOR algorithm, and these single-sided algorithms are used
primarily to monitor the stability of the LUCID detector.
There is also a LUCID\_HitOR hit counting algorithm which has been employed in the 2011 
running to cross-check the linearity of the event counting algorithms at 
high values of $\mu_{\mathrm{vis}}$.

For the BCM detector, there are two independent readout systems (BCMH and BCMV).
A hit is defined as a single sensor with a response above the noise threshold.  
Inclusive OR and coincidence AND algorithms are defined for each of these independent 
readout systems, for a total of four BCM algorithms.

\subsection{Offline algorithms}
\label{sec:offlinealg}

Additional offline analyses have been performed which rely on the MBTS and the 
vertexing capabilities of the Inner Detector.
These offline algorithms use data triggered and read out through the standard ATLAS data
acquisition system, and do not have the necessary rate capability to measure luminosity independently
for each BCID under normal physics conditions. 
Instead, these algorithms are typically used as cross-checks of the primary online algorithms
under special running conditions, where the trigger rates for these algorithms can be increased.

The MBTS system is used for luminosity measurements only for the data collected in 
the 2010 run before 150~ns bunch train operation began.
Events are triggered by the L1\_MBTS\_1\ trigger which requires at least one hit in any of the 32 MBTS counters (which is equivalent to an inclusive MBTS\_EventOR requirement).
In addition to the trigger requirement, the MBTS\_Timing analysis uses the time
measurement of the MBTS detectors to select events where the time difference between the average hit times on the two 
sides of the MBTS satisfies $|\Delta t| < 10$~ns.
This requirement is effective in rejecting beam-induced background events, as the particles produced in these events tend to traverse the detector longitudinally resulting in large values of 
$|\Delta t|$, while particles coming from the interaction point produce values of $|\Delta t|\simeq 0$.  
To form a $\Delta t$ value requires at least one hit on both sides of the IP, and so 
the MBTS\_Timing algorithm is in fact a coincidence algorithm.

Additional algorithms have been developed which are based on reconstructing interaction 
vertices formed by tracks measured in the Inner Detector. 
In 2010, the events were triggered by the L1\_MBTS\_1 trigger. 
The 2010 algorithm counts events with at least one reconstructed vertex, with at least two tracks with $p_\mathrm{T} > 100$~MeV. 
This ``primary vertex event counting'' (PrimVtx) algorithm is fundamentally an inclusive 
event-counting algorithm, and the conversion from the observed event rate to
$\mu_{\mathrm{vis}}$ follows Eq.~(\ref{eqn:muor}).

The 2011 vertexing algorithm uses events from a trigger which 
randomly selects crossings from filled bunch pairs where collisions are possible. 
The average number of visible interactions per bunch crossing is determined by counting the
number of reconstructed vertices found in each bunch crossing (Vertex). 
The vertex selection criteria in 2011 were changed to require five tracks with 
$p_\mathrm{T} > 400$~MeV while also requiring tracks to have a hit in any active pixel detector module 
along their path.

Vertex counting suffers from nonlinear behaviour with increasing interaction rates per 
bunch crossing, primarily due to two effects: vertex masking and fake vertices.
Vertex masking occurs when the vertex reconstruction algorithm fails to resolve nearby 
vertices from separate interactions, decreasing the vertex reconstruction efficiency as 
the interaction rate increases. 
A data-driven correction is derived from the distribution of distances in the longitudinal 
direction ($\Delta z$) between pairs of reconstructed vertices. 
The measured distribution of longitudinal positions ($z$) is used to predict the expected $\Delta z$
distribution of pairs of vertices if no masking effect was present.
Then, the difference between the expected and observed $\Delta z$ distributions is related to 
the number of vertices lost due to masking. The procedure is checked with
simulation for self-consistency at the sub-percent level, and the magnitude of the correction
reaches up to $+50\%$ over the range of pile-up values in 2011 physics data.
Fake vertices result from a vertex that would normally fail the requirement on the minimum number of tracks, but additional tracks from a 
second nearby interaction are erroneously assigned so that the resulting reconstructed vertex satisfies the selection criteria.
A correction is derived from simulation and reaches $-10\%$ in 2011.
Since the 2010 PrimVtx algorithm requirements are already satisfied with one reconstructed 
vertex, vertex masking has no effect, although a correction must still be made for fake vertices.

\subsection{Calorimeter-based algorithms}
\label{sec:particlealg}

The TileCal and FCal luminosity determinations do not depend upon event counting, 
but rather upon measuring detector currents that are proportional to the total particle flux in 
specific regions of the calorimeters.
These particle counting algorithms are expected to be free from pile-up effects
up to the highest interaction rates observed in late 2011 $(\mu \simeq 20)$.

The Tile luminosity algorithm measures PMT currents for selected cells in a region near 
$|\eta| \approx 1.25$ where the largest variations in current as a function of the luminosity are
observed.
In 2010, the response of a common set of cells was calibrated with respect to the luminosity 
measured by the LUCID\_EventOR algorithm in a single ATLAS run. 
At the higher luminosities encountered in 2011, 
TileCal started to suffer from frequent trips of the low-voltage power supplies, 
causing the intermittent loss of current measurements from several modules.
For these data, a second method is applied, based on the calibration of individual cells,
which has the advantage of allowing different sets of cells to be used depending on their
availability at a given time.
The calibration is performed by comparing the luminosity measured by the LUCID\_EventOR 
algorithm to the individual cell currents at the peaks of the 2011 {\em vdM} scan, as more fully described
in Sect.~\ref{sec:tilecalibration}.
While TileCal does not provide an independent absolute luminosity measurement, it
enables systematic uncertainties associated with both long-term stability and
$\mu$-dependence to be evaluated.

Similarly, the FCal high-voltage currents cannot be directly calibrated during a 
{\em vdM} scan because the total luminosity delivered in these scans remains below the sensitivity of 
the current-measurement technique.
Instead, calibrations were evaluated for each usable HV line independently by comparing to 
the LUCID\_EventOR luminosity for a single ATLAS run in each of 2010 and 2011.
As a result, the FCal also does not provide an independently calibrated luminosity measurement, 
but it can be used as a systematic check of the stability and linearity of other algorithms.
For both the TileCal and FCal analyses, the luminosity is assumed to be linearly proportional to 
the observed currents after correcting for pedestals and non-collision backgrounds.

\section{Luminosity calibration}
\label{sec:calibration}
In order to use the measured interaction rate $\mu_\mathrm{vis}$ as a luminosity monitor, 
each detector and algorithm must be calibrated by determining its visible cross-section 
$\sigma_\mathrm{vis}$.
The primary calibration technique to determine the absolute luminosity scale of each 
luminosity detector and algorithm employs dedicated {\em vdM} scans to infer the delivered 
luminosity at one point in time from the measurable parameters of the colliding bunches.
By comparing the known luminosity delivered in the {\em vdM} scan to the visible interaction rate
$\mu_\mathrm{vis}$, the visible cross-section can be determined from Eq.~(\ref{eqn:defmu}).

To achieve the desired accuracy on the absolute luminosity, these scans are not performed 
during normal physics operations, but rather under carefully controlled 
conditions with a limited number of colliding bunches and a modest peak interaction 
rate $(\mu \lesssim 2)$.
At $\sqrt s = 7$~\TeV\, three sets of such scans were performed in 2010 and one set in 2011.
This section describes the {\em vdM} scan procedure, while Sect.~\ref{sec:errors} 
discusses the systematic uncertainties on this procedure and summarizes 
the calibration results.

\subsection{Absolute luminosity from beam parameters}
\label{subsec:vdMform}

In terms of colliding-beam parameters, the luminosity $\mathcal L$ is defined 
(for beams colliding with zero crossing angle) as
\begin{equation}
{\mathcal L} = n_{\mathrm b} f_{\mathrm r} n_1 n_2 \int {\hat{\rho} _1 (x,y)} \hat{\rho} _2
(x,y)dxdy\label{lumi}
\end{equation}
where $n_{\mathrm b}$ is the number of colliding bunch pairs, 
$f_{\mathrm r}$ is the machine revolution 
frequency ($11245.5$~Hz for the LHC), $n_1 n_2$ is the bunch population product, 
and $\hat{\rho}_{1(2)}(x,y)$ is the normalized particle density 
in the transverse ($x$-$y$) plane of beam 1 (2) at the IP. 
Under the general assumption that the particle densities can be factorized
into independent horizontal and vertical components,
($\hat{\rho}(x,y)=\rho_x(x)\rho_y(y)$), Eq.~(\ref{lumi}) can be rewritten as
\begin{equation}
{\mathcal L} = n_{\mathrm b} f_{\mathrm r} n_1 n_2 
               ~\Omega _x (\rho _{x1},\rho _{x2})
               ~\Omega _y (\rho _{y1},\rho _{y2})
\label{lumi1}
\end{equation}
where \[ \Omega _x (\rho _{x1},\rho _{x2} ) = \int {\rho _{x1} (x)}\rho _{x2} (x)dx\] 
is the beam-overlap integral in the $x$ direction (with an analogous
definition in the $y$ direction). 
In the method proposed by van der Meer~\cite{bib:vdm} the overlap integral (for example in the $x$ direction) 
can be calculated as
\begin{equation}
\Omega _x (\rho _{x1},\rho _{x2}) = \frac{{R_x (0)}}{{\int {R_x (\delta)d\delta} }},\label{vdm}
\end{equation}
where $R_x(\delta)$ is the luminosity (or equivalently $\mu_\mathrm{vis}$) --- at this stage in 
arbitrary units --- measured during a horizontal scan at the time the two beams are 
separated by the distance $\delta$, and $\delta=0$ represents the 
case of zero beam separation. 

Defining the parameter $\Sigma_x$ as
\begin{equation}
\Sigma _x  = \frac{1}{{\sqrt {2\pi } }}\frac{{\int {R_x (\delta)d\delta}
}}{{R_x (0)}},
\label{caps}
\end{equation}
and similarly for $\Sigma_y$, the luminosity in Eq.~(\ref{lumi1}) can be rewritten as
\begin{equation}
{\mathcal L} = \frac{{n_{\mathrm b} f_{\mathrm r} n_1 n_2 }}{{2\pi \Sigma _x \Sigma _y
}},\label{eqn:lumifin}
\end{equation}
which enables the luminosity to be extracted from machine parameters by 
performing a {\em vdM} (beam-separation) scan. 
In the case where the luminosity curve $R_x(\delta)$ is Gaussian, 
$\Sigma _x $ coincides with the standard deviation of that distribution.
Equation~(\ref{eqn:lumifin}) is quite general; $\Sigma_x$ and $\Sigma_y$, as defined in 
Eq.~(\ref{caps}), depend only upon the area under the luminosity curve, 
and make no assumption as to the shape of that curve.

\subsection{{\em vdM} scan calibration}

To calibrate a given luminosity algorithm, one can equate the absolute luminosity computed
using Eq.~(\ref{eqn:lumifin}) to the luminosity measured by a particular algorithm at the
peak of the scan curve using Eq.~(\ref{eqn:defmu}) to get
\begin{equation}
\sigma_{\mathrm{vis}} =\mu^{\mathrm{MAX}}_{\mathrm{vis}} \frac{2\pi \Sigma_x \Sigma_y}{n_1 n_2},
\label{eqn:sigmaVis}
\end{equation}
where $\mu^{\mathrm{MAX}}_{\mathrm{vis}}$ is the visible interaction rate per bunch crossing 
observed at the peak of the scan curve as measured by that particular algorithm.
Equation~(\ref{eqn:sigmaVis}) provides a direct calibration of the visible cross-section 
$\sigma_\mathrm{vis}$ for each algorithm in terms of the peak visible interaction rate 
$\mu^{\mathrm{MAX}}_{\mathrm{vis}}$, the product of the convolved beam widths $\Sigma_x \Sigma_y$, and the bunch population product $n_1 n_2$.
As discussed below, the bunch population product must be determined from an external analysis 
of the LHC beam currents, but the remaining parameters are extracted directly from the analysis
of the {\em vdM} scan data.

For scans performed with a crossing angle, where the beams no longer collide head-on,
the formalism  becomes considerably
more involved~\cite{bib:Lconcept}, but the conclusions remain unaltered and 
Eqs.~(\ref{caps})--(\ref{eqn:sigmaVis}) remain valid. 
The non-zero vertical crossing angle used for some scans widens 
the luminosity curve by a factor that depends on the
bunch length, the transverse beam size and the crossing angle, but reduces the 
peak luminosity by the same factor. 
The corresponding increase in the measured value of $\Sigma_y$ is exactly cancelled by the decrease in $\mu^{\mathrm{MAX}}_{\mathrm{vis}}$,
 so that no correction for the crossing angle is needed in the determination of 
 $\sigma_\mathrm{vis}$.

One useful quantity that can be extracted from the {\it vdM} scan data for each luminosity 
method and that depends only on the transverse beam sizes, 
is the specific luminosity  ${\cal L}_{\mathrm{spec}}$:
\begin{equation}
{\cal L}_{\mathrm{spec}} = {\cal L}/(n_{\mathrm{b}} n_1 n_2) = \frac{f_{\mathrm{r}}}{2\pi \Sigma_x \Sigma_y}.
\label{eqn:specLumi}
\end{equation}
Comparing the specific luminosity values ({\it i.e.} the inverse product of the convolved beam sizes) measured in the same scan by different detectors and algorithms provides a direct check on the mutual consistency of the absolute luminosity scale provided by these methods.

\subsection{{\em vdM} scan data sets}
\label{sec:scanDataSets}
The beam conditions during the dedicated {\em vdM} scans are different from the conditions
in normal physics fills, with fewer bunches colliding, no bunch trains, and lower 
bunch intensities.
These conditions are chosen to reduce various systematic uncertainties in the scan procedure.

A total of five {\em vdM} scans were performed in 2010, on three different dates separated 
by weeks or months, and an additional two {\em vdM} scans at $\sqrt s = 7$~\TeV\ 
were performed in 2011 on the same day to calibrate the absolute luminosity scale.
As shown in Table~\ref{tab:vdmScan}, the scan parameters evolved from the early 2010 scans
where single bunches and very low bunch charges were used.  
The final set of scans in 2010 and the scans in 2011 were more similar, as both 
used close-to-nominal 
bunch charges, more than one bunch colliding, and typical peak $\mu$ values in the 
range 1.3--2.3.

Generally, each {\em vdM} scan consists of two separate beam scans, one where the beams
are separated by up to $\pm 6 \sigma_\mathrm{b}$ in the $x$ direction keeping the beams centred in $y$, 
and a second where the beams are separated in the $y$ direction with the beams centred in $x$, where 
$\sigma_\mathrm{b}$ is the transverse size of a single beam.
The beams are moved in a certain number of scan steps, then data are recorded for 20--30 seconds
at each step to obtain a statistically significant measurement in each luminosity detector under
calibration.
To help assess experimental systematic uncertainties in the calibration procedure, 
two sets of identical
{\em vdM} scans are usually taken in short succession to provide two 
independent calibrations under similar beam conditions.
In 2011, a third scan was performed with the beams separated by $160\, \mu$m in the 
non-scanning plane to constrain systematic uncertainties on the factorization assumption 
as discussed in Sect.~\ref{sec:correlations}.

Since the luminosity can be different for each colliding bunch pair, both because the beam 
sizes can vary bunch-to-bunch but also because the bunch population product $n_1 n_2$ can vary 
at the level of 10--20\%, the determination of $\Sigma_{x/y}$ and the measurement of 
$\mu^{\mathrm{MAX}}_{\mathrm{vis}}$ at the scan peak must be performed independently for 
each colliding BCID.
As a result, the May 2011 scan provides 14 independent measurements of 
$\sigma_{\mathrm{vis}}$ within the same scan, and the October 2010 scan provides 6.
The agreement among the $\sigma_{\mathrm{vis}}$ values extracted from these different BCIDs
provides an additional consistency check for the calibration procedure.  

\begin{table*}
\centering
\caption{Summary of the main characteristics of the 2010 and 2011 {\it vdM} scans performed at the ATLAS interaction point. 
Scan directions are indicated by ``H'' for horizontal and ``V'' for vertical.
The values of luminosity/bunch and $\mu$ are given for zero beam separation.}
\label{tab:vdmScan}
\begin{tabular*}{\textwidth}{@{\extracolsep{\fill}}lcccc@{}}
\hline
Scan Number & I & II--III & IV--V   & VII--IX \\
LHC Fill Number & 1059 & 1089 & 1386 & 1783 \\
Date  & 26 Apr., 2010    & 9 May, 2010 & 1 Oct., 2010 & 15 May, 2011\\
\hline
Scan Directions & 1 H scan           & 2 H scans  & 2 sets of & 3 sets of \\
               & followed by & followed by & H plus V scans & H plus V scans \\
               & 1 V scan & 2 V scans & & (scan IX offset) \\
Total Scan Steps per Plane &  27 & 27 & 25 & 25 \\
                           & $(\pm 6 \sigma_{\mathrm{b}})$ & $(\pm 6 \sigma_{\mathrm{b}})$ 
                           & $(\pm 6 \sigma_{\mathrm{b}})$ & $(\pm 6\sigma_{\mathrm{b}})$ \\
  Scan Duration per Step & 30~s & 30~s  & 20~s & 20~s\\
  \hline
 Bunches colliding in ATLAS \& CMS &  1  & 1 &  6 & 14 \\
 Total number of bunches per beam  & 2 & 2  & 19 & 38 \\
  Typical number of protons per bunch $(\times10^{11})$
                      & $0.1$ & $0.2$  &  $0.9$ & $0.8$\\
   Nominal $\beta$-function at IP
    [$\beta^\star$] (m)               &   $2$   &  $2$   & $3.5$ & $1.5$ \\
    Approx. transverse single beam size $\sigma_{\mathrm{b}}$ ($\mu$m) &  $45$   & $45 $& $57$ & $40$\\
    Nominal half crossing angle ($\mu$rad)    &  0    & 0 &  $\pm 100$ & $\pm 120$ \\
    \hline
    Typical luminosity/bunch ($\mu{\rm b}^{-1}/\mathrm{s})$ & $  4.5\cdot10^{-3}$ 
                                                   & $  1.8\cdot10^{-2}$ & $0.22$ & $0.38$ \\
    $\mu$ (interactions/crossing)  & 0.03 & 0.11 & 1.3 & 2.3 \\
\hline
\end{tabular*}
\end{table*}

\subsection{{\em vdM} scan analysis}
\label{sec:vdmAnalysis}

\begin{figure} 
   \centering
   \includegraphics[width=\columnwidth]{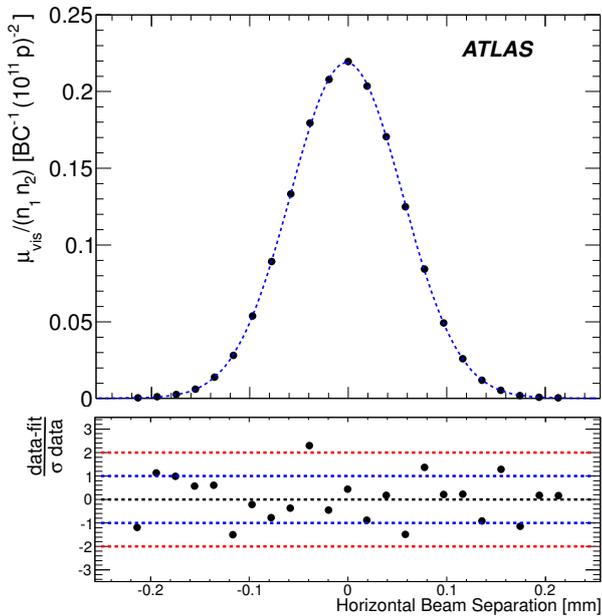} 
   \caption{Specific visible interaction rate versus nominal beam separation for the BCMH\_EventOR algorithm during scan VII in the horizontal plane for BCID 817.  
   The residual deviation of the data from the Gaussian plus constant term fit, normalized at each point to the statistical uncertainty $(\sigma\, {\mathrm{data}})$, is shown in the bottom panel.}
   \label{fig:BCMData}
\end{figure}

For each algorithm being calibrated, the {\em vdM} scan data are analysed in a very 
similar manner.
For each BCID, the specific visible interaction rate $\mu_{\mathrm{vis}}/(n_1 n_2)$ 
is measured as a function of the ``nominal'' beam separation, 
{\it i.e.} the separation specified by the LHC control system for each scan step.
The specific interaction rate is used so that the result is not affected by the change in beam 
currents over the duration of the scan.
An example of the {\em vdM} scan data for a single BCID from scan VII in the horizontal plane is shown in Fig.~\ref{fig:BCMData}.  

The value of $\mu_{\mathrm{vis}}$ is determined from the raw event rate using the 
analytic function described in Sect.~\ref{sec:mu} for the inclusive EventOR algorithms.  
The coincidence EventAND algorithms are more involved, and a numerical inversion is performed to determine
$\mu_{\mathrm{vis}}$ from the raw EventAND rate.
Since the EventAND $\mu$ determination depends on $\sigma_{\mathrm{vis}}^{\mathrm{AND}}$ as well as 
$\sigma_{\mathrm{vis}}^{\mathrm{OR}}$, an iterative procedure must be employed. 
This procedure is found to converge after a few steps.

At each scan step, the beam separation and the visible interaction rate are corrected for beam--beam effects as described in Sect.~\ref{subsec:beambeamCrctns}.  These corrected data for each BCID of each scan are then fitted independently to a characteristic function to provide a measurement of  $\mu^{\mathrm{MAX}}_{\mathrm{vis}}$ from the peak of the fitted function, while $\Sigma$ is computed from the integral of the function, using Eq.~(\ref{caps}).
Depending upon the beam conditions, this function can be a double Gaussian plus
a constant term, a single Gaussian plus a constant term, a spline function,
or other variations.
As described in Sect.~\ref{sec:errors}, the differences between the different treatments are 
taken into account as a systematic uncertainty in the calibration result.

\begin{figure} 
   \centering
   \includegraphics[width=\columnwidth]{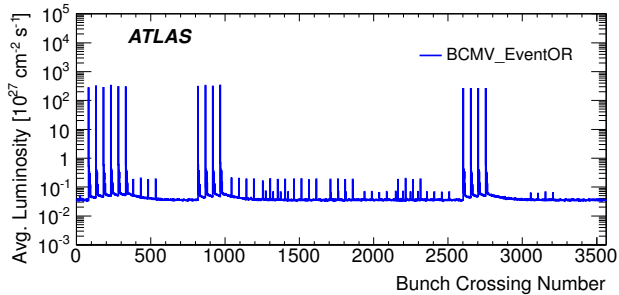} 
   \caption{Average observed luminosity per BCID from BCMV\_EventOR in the May 2011 {\em vdM} scan.  
   In addition to the 14 large spikes in the BCIDs where two bunches are colliding, induced ``afterglow'' activity can also be seen in the following BCIDs.  
   Single-beam background signals are also observed in BCIDs corresponding to unpaired bunches (24 in each beam).}
   \label{fig:backgrounds}
\end{figure}

One important difference in the {\em vdM} scan analysis between 2010 and 2011 is the treatment of the backgrounds in the luminosity signals.  
Figure~\ref{fig:backgrounds} shows the average BCMV\_EventOR luminosity as a function of BCID during the May 2011 {\em vdM} scan.  
The 14 large spikes around ${\cal L} \simeq 3\times 10^{29} \mathrm{cm}^{-2}\mathrm{s}^{-1}$ are the BCIDs containing colliding bunches.
Both the LUCID and BCM detectors observe some small activity in 
the BCIDs immediately following a collision which tends to die away to some baseline value with several different time constants.
This ``afterglow''  is most likely caused by photons from nuclear de-excitation, which in
turn is induced by the hadronic cascades initiated by $pp$ collision products.
The level of the afterglow background is observed to be proportional to the luminosity in the colliding BCIDs, and in the {\em vdM} scans 
this background can be estimated by looking at the luminosity signal in the BCID immediately preceding a colliding bunch pair.
A second background contribution comes from activity correlated with the passage of a single beam through the detector.
This ``single-beam'' background, seen in Fig.~\ref{fig:backgrounds} as the numerous small spikes at the $10^{26} \mathrm{cm}^{-2}\mathrm{s}^{-1}$ 
level, is likely a combination of beam-gas interactions and halo particles which intercept the luminosity detectors in time with the main beam.
It is observed that this single-beam background is proportional to the bunch charge present in each bunch, and can be considerably different 
for beams 1 and 2, but is otherwise uniform for all bunches in a given beam.
The single-beam background underlying a collision BCID can be estimated by measuring the single-beam backgrounds in unpaired bunches 
and correcting for the difference in bunch charge between the unpaired and colliding bunches. 
Adding the single-beam backgrounds measured for beams 1 and 2 then gives an estimate for the single-beam background present in 
a colliding BCID.
Because the single-beam background does not depend on the luminosity, this background can dominate the observed luminosity response when 
the beams are separated.

In 2010, these background sources were accounted for by assuming that any constant term fitted to the observed scan curve is the result of 
luminosity-independent background sources, and has not been included as part of the luminosity integrated to extract $\Sigma_x$ or $\Sigma_y$.  
In 2011, a more detailed background subtraction is first performed to correct each BCID for afterglow and single-beam backgrounds, then 
any remaining constant term observed in the scan curve has been treated as a broad luminosity signal which contributes to the determination of $\Sigma$.

The combination of one $x$ scan and one $y$ scan is the minimum needed 
to perform a measurement of $\sigma_{\mathrm{vis}}$.
The average value of  $\mu^{\mathrm{MAX}}_{\mathrm{vis}}$ between the two scan planes is 
used in the determination of $\sigma_{\mathrm{vis}}$, and the correlation matrix from each fit 
between $\mu^{\mathrm{MAX}}_{\mathrm{vis}}$ and $\Sigma$ is taken into account when 
evaluating the statistical uncertainty.

Each BCID should measure the same $\sigma_{\mathrm{vis}}$ value, and the average over all BCIDs is taken as the $\sigma_{\mathrm{vis}}$ measurement for that scan.  
Any variation in $\sigma_{\mathrm{vis}}$ between BCIDs, as well as between scans, 
reflects the reproducibility and stability of the calibration procedure during a single fill.

\begin{figure} 
   \centering
   \includegraphics[width=\columnwidth]{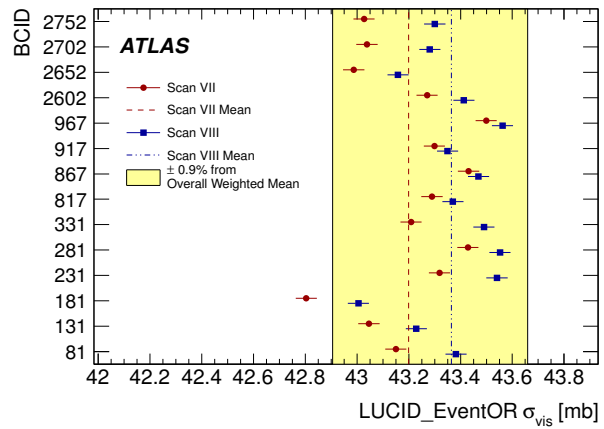} 
   \caption{Measured $\sigma_{\mathrm{vis}}$ values for LUCID\_EventOR by BCID for scans VII and VIII.  The error bars represent statistical errors only.  The vertical lines indicate the weighted average over BCIDs for scans VII and VIII separately. The shaded band indicates a  $\pm 0.9\%$ variation from the average, which is the systematic uncertainty evaluated from the per-BCID and per-scan $\sigma_{\mathrm{vis}}$ consistency.
}
   \label{fig:SigmaVisLUCIDMay}
\end{figure}

\begin{figure} 
   \centering
   \includegraphics[width=\columnwidth]{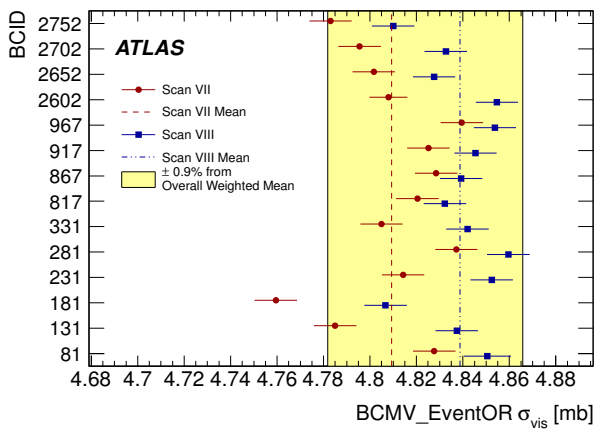} 
   \caption{Measured $\sigma_{\mathrm{vis}}$ values for BCMV\_EventOR by BCID for scans VII and VIII.  The error bars represent statistical errors only.  The vertical lines indicate the weighted average over BCIDs for Scans VII and VIII separately. The shaded band indicates a  $\pm 0.9\%$ variation from the average, which is the systematic uncertainty evaluated from the per-BCID and per-scan $\sigma_{\mathrm{vis}}$ consistency.  
}
   \label{fig:SigmaVisBCMMay}
\end{figure}

Figure~\ref{fig:SigmaVisLUCIDMay} shows the $\sigma_{\mathrm{vis}}$ values determined for LUCID\_EventOR separately by BCID and by scan in the May 2011 scans.
The RMS variation seen between the $\sigma_{\mathrm{vis}}$ results measured for different BCIDs is $0.4\%$ for scan VII and $0.3\%$ for scan VIII.
The BCID-averaged $\sigma_{\mathrm{vis}}$ values found in scans VII and VIII agree to $0.5\%$ (or better) for all four LUCID algorithms.
Similar data for the BCMV\_EventOR algorithm are shown in Fig.~\ref{fig:SigmaVisBCMMay}.
Again an RMS variation between BCIDs of up to $0.55\%$ is seen, and a difference between the two 
scans of up to $0.67\%$ is observed for the BCM\_EventOR algorithms.
The agreement in the BCM\_EventAND algorithms is worse, with an RMS around $1\%$, although these measurements also have significantly larger statistical errors.

Similar features are observed in the October 2010 scan, where the $\sigma_\mathrm{vis}$ results 
measured for different BCIDs, and the BCID-averaged $\sigma_\mathrm{vis}$ value found in scans IV 
and V agree to 0.3\% for LUCID\_EventOR and 0.2\% for LUCID\_EventAND.
The BCMH\_EventOR results agree between BCIDs and between the two scans at the 0.4\% level,
while the BCMH\_EventAND calibration results are consistent within the larger statistical errors 
present in this measurement.

\subsection{Internal scan consistency}
\label{sec:scanConsistency}

The variation between the measured $\sigma_{\mathrm{vis}}$ values by BCID and between scans 
quantifies the stability and reproducibility of the calibration technique.
Comparing Figs.~\ref{fig:SigmaVisLUCIDMay} and \ref{fig:SigmaVisBCMMay} for the May 2011
scans, it is clear that some of the variation seen in $\sigma_{\mathrm{vis}}$ 
is not statistical in nature, but rather is correlated by BCID.
As discussed in Sect.~\ref{sec:errors}, the RMS variation of $\sigma_{\mathrm{vis}}$ 
between BCIDs within a given scan is taken as a systematic uncertainty in the 
calibration technique, 
as is the reproducibility of $\sigma_{\mathrm{vis}}$ between scans.
The yellow band in these figures, which represents a range of $\pm 0.9\%$, 
shows the quadrature sum of these two systematic uncertainties.
Similar results are found in the final scans taken in 2010, although with only 6 
colliding bunch pairs there are fewer independent measurements to compare.

\begin{figure} 
   \centering
   \subfigure{\includegraphics[width=\columnwidth]{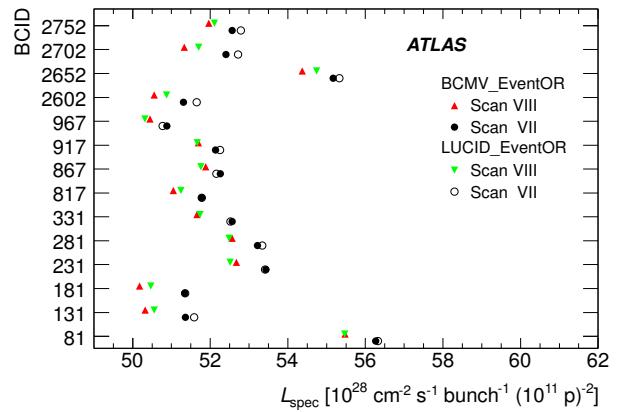}}
    \subfigure{\includegraphics[width=\columnwidth]{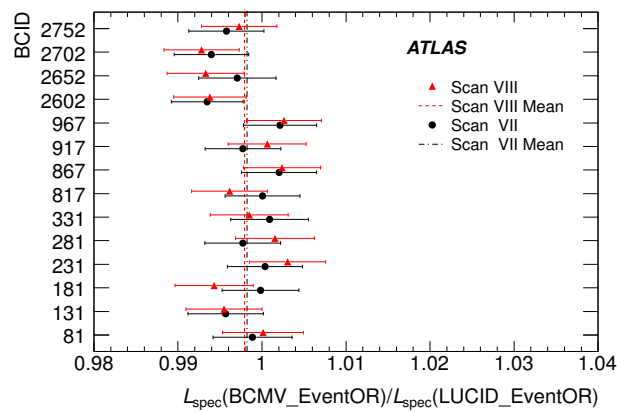}}
   \caption{Specific luminosity determined by BCMV and
    LUCID per BCID for scans VII and VIII.  
   The figure on the top shows the specific luminosity values determined by  BCMV\_EventOR 
   and LUCID\_EventOR, 
   while the figure on the bottom shows the ratios of these values.
   The vertical lines indicate the weighted average over BCIDs for scans VII and VIII separately.
   The error bars represent statistical uncertainties only.
}

   \label{fig:LSpec}
\end{figure}

Further checks can be made by considering the distribution of ${\cal L}_{\mathrm{\mathrm{spec}}}$ 
defined in Eq.~(\ref{eqn:specLumi}) for a given BCID as measured by different algorithms.
Since this quantity depends only on the convolved beam sizes, consistent results should be
measured by all methods for a given scan.
Figure~\ref{fig:LSpec} shows the measured  ${\cal L}_{\mathrm{\mathrm{spec}}}$ values by 
BCID and scan for LUCID and BCMV algorithms, as well as the ratio of these values in 
the May 2011 scans.
Bunch-to-bunch variations of the specific luminosity are typically 5--10\%, reflecting 
bunch-to-bunch differences in transverse emittance also seen during normal physics fills.
For each BCID, however, all algorithms are statistically consistent.
A small systematic reduction in ${\cal L}_{\mathrm{spec}}$ can be observed between 
scans VII and VIII, which is due to emittance growth in the colliding beams.

\begin{figure} 
   \centering
   \subfigure{\includegraphics[width=\columnwidth]{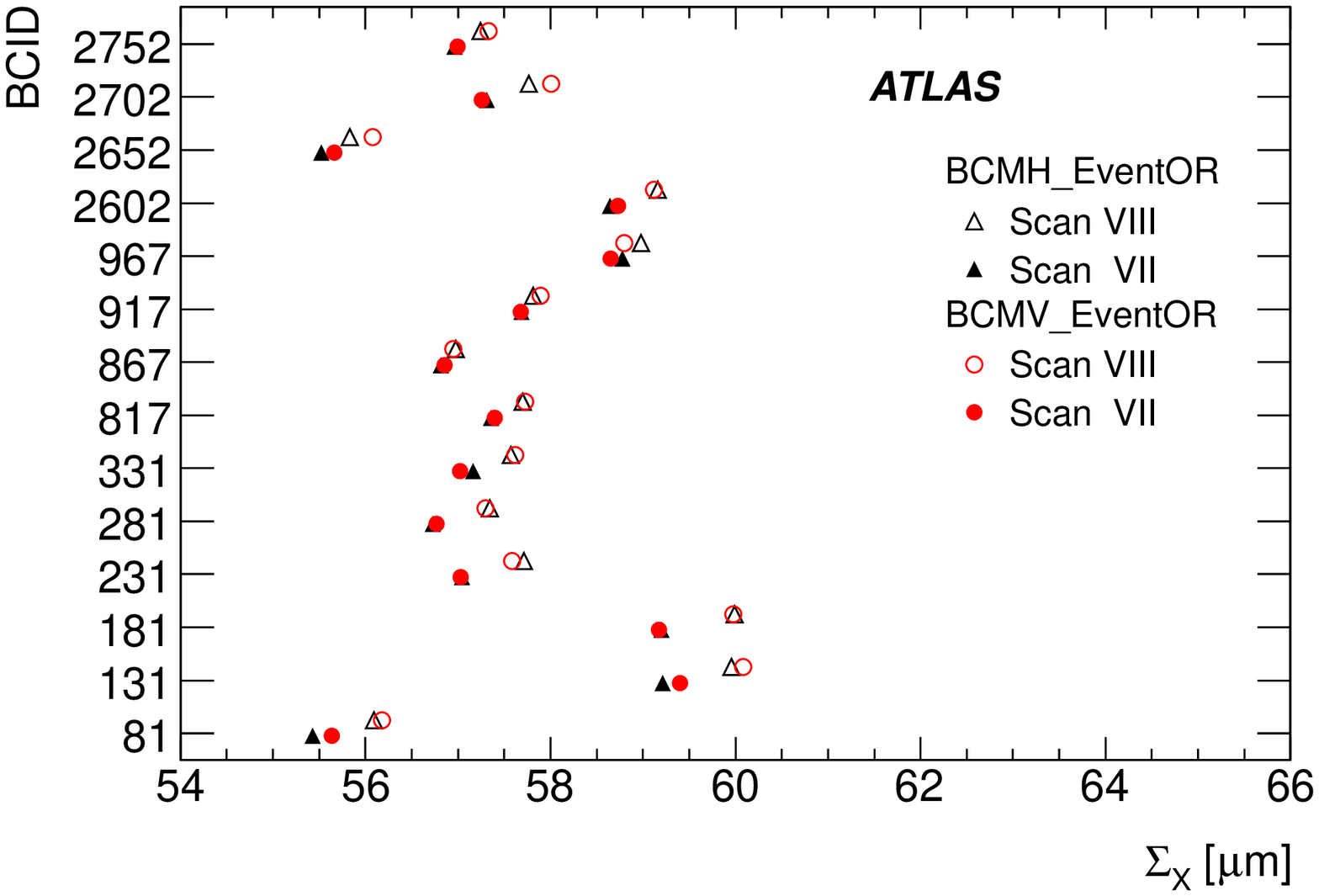}}
   \caption{$\Sigma_x$ determined by BCM\_EventOR algorithms per BCID for scans VII and VIII.  The statistical uncertainty on each measurement is approximately the size of the marker.
}
 \label{fig:CapSigmaX}
\end{figure}

\begin{figure} 
   \centering
   \subfigure{\includegraphics[width=\columnwidth]{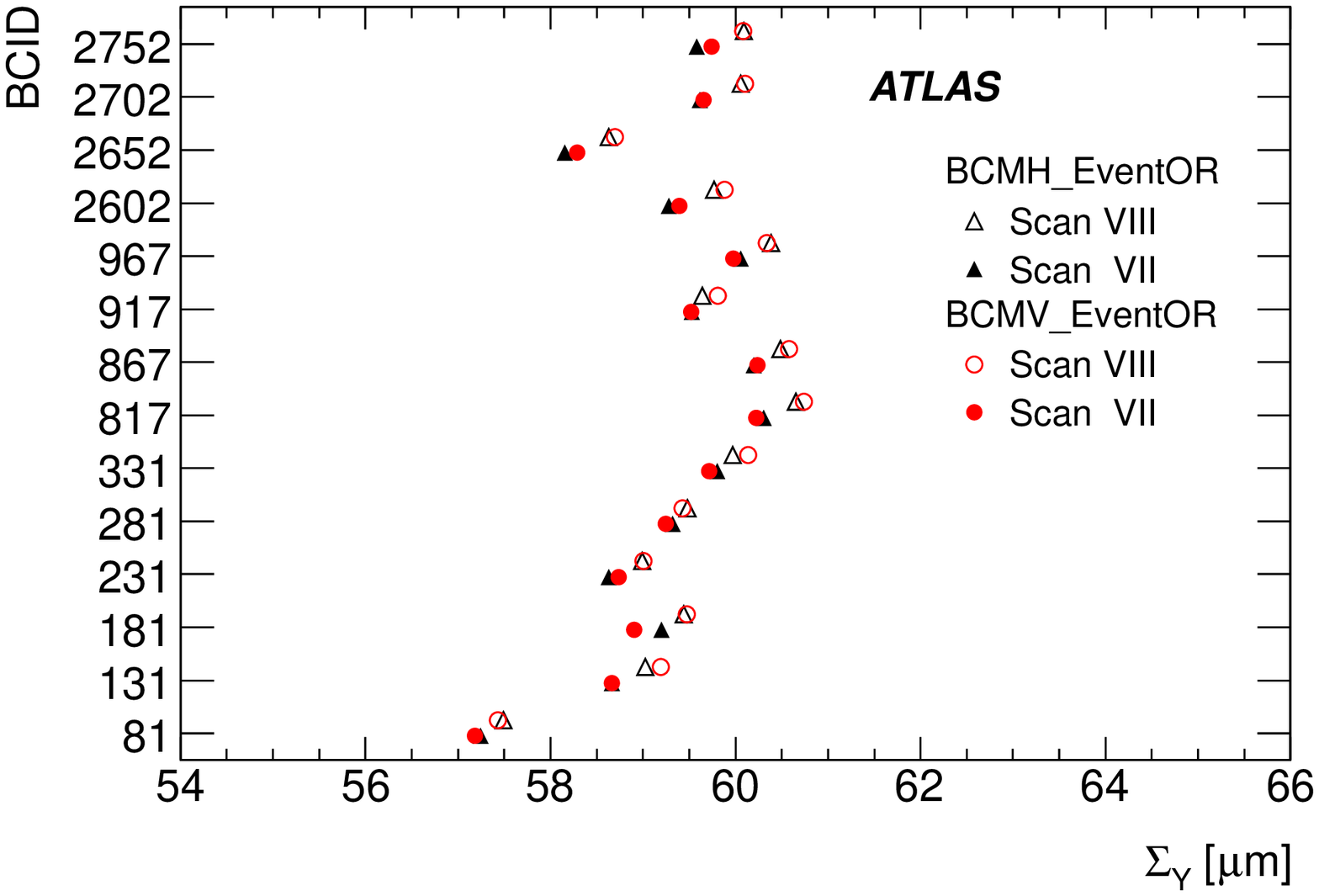}} 
   \caption{$\Sigma_y$ determined by BCM\_EventOR algorithms per BCID for scans VII and VIII.  The statistical uncertainty on each measurement is approximately the size of the marker.
}
 \label{fig:CapSigmaY}
\end{figure}

\begin{figure} 
   \centering
   \includegraphics[width=\columnwidth]{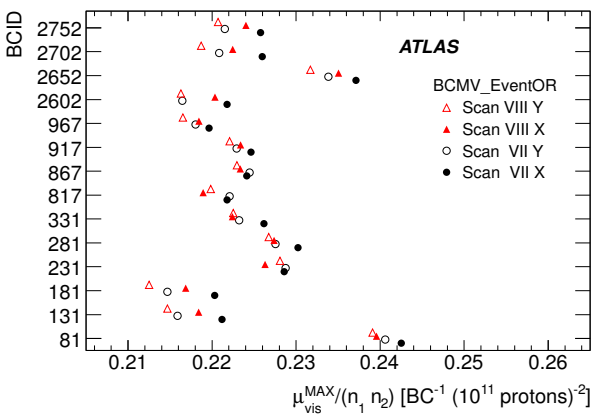}
   \caption{Peak specific interaction rate $\mu^{\mathrm{MAX}}_{\mathrm{vis}}/(n_1 n_2)$ determined by BCMV\_EventOR per BCID for scans VII and VIII.   The statistical uncertainty on each measurement is approximately the size of the marker.
}
 \label{fig:muVis}
\end{figure}

Figures~\ref{fig:CapSigmaX} and \ref{fig:CapSigmaY} show the $\Sigma_x$ and $\Sigma_y$ values determined by the BCM algorithms 
during scans VII and VIII, and for each BCID a clear increase can be seen with time.
This emittance growth can also be seen clearly as a reduction in the peak specific interaction rate $\mu_{\mathrm{vis}}^{\mathrm{MAX}}/(n_1 n_2)$ shown in Fig.~\ref{fig:muVis} for BCMV\_EventOR.
Here the peak rate is shown for each of the four individual horizontal and vertical scans, and a  monotonic decrease in rate is generally observed as each individual scan curve is recorded.
The fact that the $\sigma_{\mathrm{vis}}$ values are consistent between scan VII and scan VIII demonstrates that to first order the emittance growth 
cancels out of the measured luminosity calibration factors.
The residual uncertainty associated with emittance growth is discussed in Sect.~\ref{sec:errors}.

\subsection{Bunch population determination}
\label{sec:currents}

The dominant systematic uncertainty on the  2010 luminosity calibration, and a significant
uncertainty  on the 2011 calibration, 
is associated with the determination of the bunch population product $(n_1 n_2)$ for 
each colliding BCID.
Since the luminosity is calibrated on a bunch-by-bunch basis for the reasons described in
Sect.~\ref{sec:scanDataSets}, the bunch population per BCID is necessary to perform this calibration.
 Measuring the bunch population product separately for each BCID is also unavoidable as only a
 subset of the circulating bunches collide in ATLAS (14 out of 38 during the 2011 scan).

The bunch population measurement is performed by the LHC 
Bunch Current Normalization Working Group (BCNWG) and has been described in detail
in Refs.~\cite{bib:BCNWG1, bib:BCNWG2} for 2010 and Refs.~\cite{bib:BCNWG3, bib:BCNWG4, bib:BCNWG5} for 2011.
A brief summary of the analysis is presented here, along with the uncertainties on the bunch
population product.
The relative uncertainty on the bunch population product ($n_1 n_2$) is shown in 
Table~\ref{tab:BCNWG} for the {\em vdM} scan fills in 2010 and 2011.

\begin{table}
   \centering
   \caption{Systematic uncertainties on the determination of the bunch population product $n_1 n_2$ for the 2010 and 2011 {\it vdM} scan fills.  The uncertainty on ghost charge and satellite bunches is included in the bunch-to-bunch fraction for scans I--V.}
   \label{tab:BCNWG}
   \begin{tabular*}{\columnwidth}{@{\extracolsep{\fill}}lcccc@{}}
      	\hline
      	Scan Number & I & II--III & IV--V &  VII--VIII \\
      	LHC Fill Number & 1059 & 1089 & 1386 & 1783 \\
      	\hline
		DCCT baseline offset  	& 3.9\% & 1.9\% & 0.1\% & 0.10\% \\
		DCCT scale variation	& 2.7\% & 2.7\% & 2.7\% & 0.21\% \\
		Bunch-to-bunch fraction	& 2.9\% & 2.9\% & 1.6\% & 0.20\%  \\
 		Ghost charge and satellites & - &  - & - & 0.44\%  \\
		\hline
		Total & 5.6\% & 4.4\% & 3.1\% & 0.54\% \\
		\hline  
   \end{tabular*}
\end{table}

The bunch currents in the LHC are determined by eight Bunch Current Transformers (BCTs) 
in a multi-step process due to the different capabilities of the available instrumentation.
Each beam is monitored by two identical and redundant DC current transformers (DCCT) 
which are high-accuracy devices but do not have any ability to separate individual bunch populations.
Each beam is also monitored by two fast beam-current transformers (FBCT) which 
have the ability to measure bunch currents individually for each of the 3564 
nominal 25 ns slots in each beam.
The relative fraction of the total current in each BCID can be determined from 
the FBCT system, but this relative measurement
must be normalized to the overall current scale provided by the DCCT.  
Additional corrections are made for any out-of-time charge that may be present in a 
given BCID but not colliding at the interaction point.

The DCCT baseline offset is the dominant uncertainty on the bunch population product in early 2010.
The DCCT is known to have baseline drifts for a variety of reasons including temperature 
effects, mechanical vibrations, and electromagnetic pick-up in cables.
For each {\em vdM} scan fill the baseline readings for each beam (corresponding to zero current) 
must be determined by looking at periods with no beam immediately before and after each fill.
Because the baseline offsets vary by at most $\pm 0.8\times 10^9$ protons 
in each beam, the relative uncertainty from the baseline determination decreases as the total
circulating currents go up. 
So while this is a significant uncertainty in scans I--III, for the remaining scans which were taken
at higher beam currents, this uncertainty is negligible.

In addition to the baseline correction, the absolute scale of the DCCT must be understood.
A precision current source with a relative accuracy of 0.1\% is used to calibrate the DCCT system
at regular intervals, and the peak-to-peak variation of the measurements made in 2010 is 
used to set an uncertainty on the bunch current product of $\pm 2.7\%$.
A considerably more detailed analysis has been performed on the 2011 DCCT data as described in Ref.~\cite{bib:BCNWG3}.
In particular, a careful evaluation of various sources of systematic uncertainties and dedicated measurements
to constrain these sources results in an uncertainty on the absolute DCCT scale in 2011 of 0.2\%.

Since the DCCT can measure only the total bunch population in each beam, 
the FBCT is used to determine the relative fraction of  bunch population in each BCID, 
such that the bunch population product colliding in a particular BCID can be determined.
To evaluate possible uncertainties in the bunch-to-bunch determination, checks are made by 
comparing the FBCT measurements to other systems which have sensitivity to the 
relative bunch population, including the ATLAS beam pick-up timing system.
As described in Ref.~\cite{bib:BCNWG4}, the agreement between the various determinations of the bunch population
is used to determine an uncertainty on the relative bunch population fraction.
This uncertainty is significantly smaller for 2011 because of a more sophisticated analysis, that exploits the consistency 
requirement that the visible cross-section be bunch-independent.

Additional corrections to the bunch-by-bunch fraction are made to correct for  ``ghost charge'' and ``satellite bunches''.
Ghost charge refers to protons that are present in nominally empty BCIDs at a level below the FBCT threshold 
(and hence invisible), but still contribute to the current measured by the more precise DCCT.
Satellite bunches describe out-of-time protons present in collision BCIDs that are measured by the FBCT, 
but that remain captured in an RF-bucket at least one period (2.5\,ns) away from the nominally filled LHC bucket, 
and as such experience only long-range encounters with the nominally filled bunches in the other beam.
These corrections, as well as the associated systematic uncertainties, 
are described in detail in Ref.~\cite{bib:BCNWG5}.

\subsection{Length scale determination}
\label{sec:LengthScale}

Another key input to the {\it vdM} scan technique is the knowledge of the beam separation at 
each scan point.
The ability to measure $\Sigma_{x/y}$ depends upon knowing the 
absolute distance by which the beams are separated during the {\em vdM} scan, which is controlled by a set of 
closed orbit bumps\footnote{A closed orbit bump is a local distortion of the beam orbit
that is implemented using pairs of steering dipoles located on either
side of the affected region. In this particular case, these bumps are tuned to offset the trajectory of
 either beam parallel to itself at the IP, in either the horizontal or the vertical direction.} 
applied locally near the ATLAS IP using steering correctors.
To determine this beam-separation length scale, dedicated length scale calibration
measurements are performed close in time to each {\em vdM} scan set using the same 
collision-optics configuration at the interaction point.
Length scale scans are performed by displacing the beams in collision by five steps over a range 
of up to $\pm 3 \sigma_\mathrm{b}$.  
Because the beams remain in collision during these scans, the actual position of the luminous 
region can be reconstructed with high accuracy using the primary vertex position reconstructed
by the ATLAS tracking detectors. 
Since each of the four bump amplitudes (two beams in two transverse directions) depends 
on different magnet and lattice functions, the distance-scale calibration scans are 
performed so that each of these four calibration constants can be extracted independently.
These scans have verified the nominal length scale assumed in the
LHC control system at the ATLAS IP at the level of $\pm 0.3$\%.

\subsection{Beam--beam corrections}
\label{subsec:beambeamCrctns}

When charged-particle bunches collide, the electromagnetic field generated by a bunch in beam 1 distorts the individual particle trajectories in the corresponding bunch of beam 2 (and vice-versa). This so-called {\em beam--beam interaction} affects the scan data in two ways.

The first phenomenon, called {\em dynamic $\beta$}~\cite{bib:dynBeta}, arises from the mutual defocusing of the two colliding bunches: this effect is tantamount to inserting a small quadrupole at the collision point. The resulting fractional change in $\beta^*$ (the value of the $\beta$ function\footnote{The $\beta$ function describes the single-particle motion and determines the variation of the beam envelope along the beam orbit. It is calculated from the focusing properties of the magnetic lattice (see for example Ref.~\cite{bib:betaDef}).}
at the IP), or equivalently the optical demagnification between the LHC arcs and the collision point, varies with the transverse beam separation,  sligthly modifying the collision rate at each scan step and thereby distorting the shape of the {\em vdM} scan curve. 

Secondly, when the bunches are not exactly centred on each other in the $x$-$y$ plane, their electromagnetic repulsion induces a mutual angular kick~\cite{bib:SLCdefl} that distorts the closed orbits by a fraction of a micrometer and modulates the actual transverse separation at the IP in a manner that depends on the separation itself. If left unaccounted for, these {\em beam--beam deflections} would bias the measurement of the overlap integrals in a manner that depends on the bunch parameters.

The amplitude and the beam-separation dependence of both effects depend similarly on the beam energy, the tunes\footnote{The tune of a storage ring is defined as the betatron phase advance per turn, or equivalently as the number of betatron oscillations over one full ring circumference.} and the unperturbed $\beta$-functions, as well as the bunch intensities and transverse beam sizes. The dynamic evolution of $\beta^*$ during the scan is modelled using the MAD-X optics code~\cite{bib:MADX} assuming bunch parameters representative of the May 2011 {\em vdM} scan (fill 1783), and then scaled using the measured intensities and convolved beam sizes of each colliding-bunch pair. The correction function is intrinsically independent of whether the bunches collide in ATLAS only, or also at other LHC interaction points~\cite{bib:dynBeta}. The largest $\beta^*$ variation during the 2011 scans is about $0.9\%$. 

The beam--beam deflections and associated orbit distortions are calculated analytically~\cite{bib:bbDefl} assuming elliptical gaussian beams that collide in ATLAS only. For a typical bunch, the peak angular kick during the 2011 scans is about $\pm 0.5 \mu$rad, and the corresponding peak increase in relative beam separation amounts to $\pm 0.6 \mu$m.  The MAD-X simulation is used to validate this analytical calculation, and to verify that higher-order dynamical effects (such as the orbit shifts induced at other collision points by beam--beam deflections at the ATLAS IP) result in negligible corrections to the analytical prediction.

At each scan step, the measured visible interaction rate is rescaled by the ratio
of the dynamic to the unperturbed bunch-size product, and the predicted change in beam separation is added to the nominal beam separation. 
Comparing the results of the scan analysis in Sect.~\ref{sec:vdmAnalysis} with and without beam--beam corrections for 
the 2011 scans, it is found that the visible cross-sections are increased by approximately 0.4\% from the 
dynamic-$\beta$ correction and 1.0\% from the deflection correction.
The two corrections combined amount to +1.4\% for 2011, and to +2.1\% for the October 2010 scans\footnote{For 2010, the correction is computed for scans IV and V only, because the bunch intensities during the earlier scans are so low as to make beam--beam effects negligible.}, reflecting the smaller emittances and slightly larger bunch intensities in that scan session.

\subsection{{\em vdM} scan results}

\begin{table}
   \centering
   \caption{Visible cross-section measurements (in mb) determined from {\it vdM} scan data in 2011.  Errors shown are statistical only.
   }
   \label{tab:sigvis2011}
   \begin{tabular*}{\columnwidth}{@{\extracolsep{\fill}}lr@{\extracolsep{0pt}$\,\pm\,$}l@{\extracolsep{\fill}}r@{\extracolsep{0pt}$\,\pm\,$}l@{\extracolsep{\fill}}@{}}
      	\hline
      	Scan Number & \multicolumn{2}{c}{VII} & \multicolumn{2}{c}{VIII}  \\
      	Fill Number & \multicolumn{2}{c}{1783} & \multicolumn{2}{c}{1783} \\
      	\hline
		LUCID\_EventAND	& 13.660 &  0.003 &  13.726 &  0.003 \\ 
		LUCID\_EventOR 	& 43.20 &  0.01 &  43.36 &  0.01 \\ 
		LUCID\_EventA        & 28.44 &  0.01 &  28.54 &  0.01 \\ 
		LUCID\_EventC        & 28.48 &  0.01 &  28.60 &  0.01 \\
		BCMH\_EventAND   & 0.1391 &  0.0004 &  0.1404 &  0.0004  \\ 
		BCMV\_EventAND   & 0.1418 &  0.0004 &  0.1430 &  0.0004  \\ 
		BCMH\_EventOR     & 4.762 &  0.002 &  4.792 &  0.003 \\ 
		BCMV\_EventOR     & 4.809 &  0.003 &  4.839 &  0.003 \\ 
		Vertex (5 tracks)     & 39.00 &  0.02 & 39.12 &  0.02 \\ 
		\hline
   \end{tabular*}
\end{table}

\begin{table*}
   \centering
   \caption{Visible cross-section measurements (in mb) determined from {\it vdM} scan data in 2010.  Errors shown are statistical only.
}
   \label{tab:sigvis2010}
   \begin{tabular*}{\textwidth}{@{\extracolsep{\fill}}lr@{\extracolsep{0pt}$\,\pm\,$}l@{\extracolsep{\fill}}r@{\extracolsep{0pt}$\,\pm\,$}l@{\extracolsep{\fill}}r@{\extracolsep{0pt}$\,\pm\,$}l@{\extracolsep{\fill}}r@{\extracolsep{0pt}$\,\pm\,$}l@{\extracolsep{\fill}}r@{\extracolsep{0pt}$\,\pm\,$}l@{}} 
      	\hline
      	Scan Number & \multicolumn{2}{c}{I} & \multicolumn{2}{c}{II} & \multicolumn{2}{c}{III} & \multicolumn{2}{c}{IV} & \multicolumn{2}{c}{V} \\
      	Fill Number & \multicolumn{2}{c}{1059} & \multicolumn{2}{c}{1089} & \multicolumn{2}{c}{1089} & \multicolumn{2}{c}{1386} & \multicolumn{2}{c}{1386} \\
      	\hline
		LUCID\_EventAND	& 11.92 &  0.14 & 12.65 &  0.10 & 12.83 &  0.10 & 13.38 &  0.01 & 13.34 &  0.01\\
		LUCID\_EventOR 	& 38.86 &  0.32 & 41.03 &  0.13 & 41.10 &  0.14 & 42.73 &  0.03 & 42.60 &  0.02\\
		BCMH\_EventAND   & \multicolumn{2}{c}{} & \multicolumn{2}{c}{}  & \multicolumn{2}{c}{} & 0.1346 &  0.0007 & 0.1341 &  0.0007 \\
		BCMH\_EventOR     & \multicolumn{2}{c}{} & \multicolumn{2}{c}{}  & \multicolumn{2}{c}{} & 4.697 &  0.007 & 4.687 &  0.007 \\
		MBTS\_Timing		& 48.3 &  0.3 & 50.2 &  0.2 & 49.9 &  0.2 & 52.4 & 0.2 & 52.3 & 0.2 \\
		PrimVtx			& 46.6 &  0.3 & 48.2 &  0.2 & 48.4 &  0.2 & 50.5 & 0.2 & 50.4 & 0.2 \\
		\hline
   \end{tabular*}
\end{table*}

The calibrated visible cross-section results for the {\em vdM} scans performed in 2010 and 2011
are shown in Tables~\ref{tab:sigvis2011} and \ref{tab:sigvis2010}. 
There were four algorithms which were calibrated in all five 2010 scans, while the
BCMH algorithms were only available in the final two scans.  
The BCMV algorithms were not considered for luminosity measurements in 2010.
Due to changes in the hardware or algorithm details between 2010 and 2011, 
the $\sigma_{\mathrm{vis}}$ values are not expected to be exactly the same in the two years.

\section{Calibration uncertainties and results}
\label{sec:errors}
This section outlines the systematic uncertainties which have been evaluated for the
measurement of  $\sigma_{\mathrm{vis}}$ from the {\it vdM} calibration scans for 2010 and 2011,
and summarizes the calibration results.
For scans~I--III, the ability to make internal cross-checks is limited due to the presence of only one colliding bunch pair in these scans, 
and the systematic uncertainties for these scans are unchanged from those evaluated in
Ref.~\cite{bib:ATLASLumiNote}.
Starting with scans~IV and V, the redundancy from having multiple bunch pairs colliding has allowed a much more detailed
study of systematic uncertainties.

The five different scans taken in 2010 have different systematic uncertainties, and the 
combination process used to determine a single $\sigma_{\mathrm{vis}}$ value is 
described in Sect.~\ref{sec:combination}.
For 2011, the two {\em vdM} scans are of equivalent quality, and the calibration results are simply 
averaged based on the statistical uncertainties.
Tables~\ref{tab:syst2010} and \ref{tab:syst2011} summarize the systematic uncertainties on the 
calibration in 2010 and 2011 respectively, while the combined calibration results are 
shown in Table~\ref{tab:sigvisresult}.

\subsection{Calibration uncertainties}

\subsubsection{Beam centring}

If the beams are not perfectly centred in the non-scanning plane at the start of a {\it vdM} scan, 
the assumption that the luminosity observed at the peak is equal to the maximum 
head-on luminosity is not correct.
In the last set of 2010 scans and the 2011 scans, the beams were centred at the beginning 
of the scan session, and the maximum observed non-reproducibility in relative beam position 
at the peak of the fitted scan curve is used to 
determine the uncertainty.  
For instance, in the 2011 scan the maximum offset is 3\,$\mu$m, corresponding 
to a $0.1\%$ error on the peak instantaneous interaction rate.

\subsubsection{Beam-position jitter}

At each step of a scan, the actual beam separation may be affected by 
random deviations of the beam positions from their nominal setting. 
The magnitude of this potential ``jitter'' has been evaluated from the 
shifts in relative beam centring recorded during the length-scale 
calibration scans described in Sect.~\ref{sec:LengthScale}, 
and amounts to aproximately 0.6~$\mu$m RMS. 
Very similar values are observed in 2010 and 2011.
The resulting systematic uncertainty on $\sigma_{\mathrm{vis}}$ is obtained by 
randomly displacing each measurement point by this amount in a series of simulated scans, 
and taking the RMS of the resulting variations in fitted visible 
cross-section. This procedure yields a $\pm 0.3$\% systematic error 
associated with beam-positioning jitter during scans IV--VIII.
For scans I--III, this is assumed to be part of the 3\% non-reproducibility uncertainty.

\subsubsection{Emittance growth}

The {\it vdM} scan formalism assumes that the luminosity and the convolved beam sizes 
$\Sigma_{x/y}$ are constant, or more precisely that the transverse 
emittances of the two beams do not vary significantly either in the 
interval between the horizontal and the associated vertical scan, or 
within a single $x$ or $y$ scan.

Emittance growth between scans would manifest itself by a slight 
increase of the measured value of $\Sigma$ from one scan to the next.
At the same time, emittance growth would decrease the peak specific luminosity in successive scans
({\it i.e.}~reduce the specific visible interaction rate at zero beam separation). 
Both effects are clearly visible in the 2011 May scan data presented in 
Sect.~\ref{sec:scanConsistency}, where
Figs.~\ref{fig:CapSigmaX} and \ref{fig:CapSigmaY} show the increase in $\Sigma$ and 
Fig.~\ref{fig:muVis} shows the reduction in the peak interaction rate.

In principle, when computing the visible cross-section using Eq.~(\ref{eqn:sigmaVis}),
the increase in $\Sigma$ from scan to scan should exactly cancel the decrease 
in specific interaction rate. 
In practice, the cancellation is almost complete: 
the bunch-averaged visible cross-sections measured in scans IV--V differ by at most $0.5\%$, 
while in scans VII--VIII the values differ by at most $0.67\%$.
These maximum differences are taken as  estimates of the systematic uncertainties due to emittance growth.

Emittance growth within a scan would manifest itself by a very slight 
distortion of the scan curve. 
The associated systematic uncertainty 
determined by a toy Monte Carlo study with the observed level of emittance growth 
was found to be negligible.

For scans I--III, an uncertainty of 3\% was determined from the variation in the peak specific interaction rate between successive scans.
This uncertainty is assumed to cover both emittance growth and other unidentified sources of non-reproducibility.  Variations of such magnitude were not observed in later scans.

\subsubsection{Consistency of bunch-by-bunch visible cross-sections}

The calibrated $\sigma_{\mathrm{vis}}$ value found for a given detector and algorithm should be a 
constant factor independent of machine conditions or BCID.
Comparing the $\sigma_{\mathrm{vis}}$ values determined by BCID in  
Figs.~\ref{fig:SigmaVisLUCIDMay} and \ref{fig:SigmaVisBCMMay}, 
however, it is clear that there is some degree of correlation between these values: the scatter observed is not entirely statistical in nature.
The RMS variation of $\sigma_{\mathrm{vis}}$ for each of the LUCID and BCM algorithms is consistently around $0.5\%$, except for the 
BCM\_EventAND algorithms, which have much larger statistical uncertainties.
An additional uncertainty of $\pm 0.55\%$ 
has been applied, corresponding to the largest RMS variation observed in either the LUCID or BCM measurements to account for this observed 
BCID dependence in 2011.
For the 2010 scans, only scans IV--V have multiple BCIDs with collisions, and in those scans the agreement between BCIDs and between scan 
sessions was consistent with the statistical accuracy of the comparison.
As such, no additional uncertainty beyond the 0.5\% derived for emittance growth was assigned.

\subsubsection{Fit model}

The {\it vdM} scan data in 2010 are analysed using a fit to a double Gaussian plus a 
constant background term, while for 2011 the data are first corrected for known backgrounds, then fitted to a single Gaussian plus constant term.
Refitting the data with several different model assumptions including a cubic spline function and no constant  term leads to different 
values of $\sigma_{\mathrm{vis}}$.
The maximum variation between these different fit assumptions is used to set an uncertainty on the fit model.

\subsubsection{Background subtraction}

The importance of the background subtraction used in the 2011 {\em vdM} analysis is evaluated by comparing the visible cross-section measured
by the BCM\_EventOR algorithms when the detailed background subtraction is performed or not performed before fitting the scan curve.
Half the difference $(0.31\%)$ is adopted as a systematic uncertainty on this procedure.
For scans IV--V, no dedicated background subtraction was performed and the uncertainty on the background treatment is accounted for in the fit 
model uncertainty, where one of the comparisons is between assuming 
the constant term results from luminosity-independent background sources compared to a luminosity-dependent signal.

\subsubsection{Reference specific luminosity}

The transverse convolved beam sizes $\Sigma_{x/y}$ measured by the {\em vdM} scan are directly related to the specific luminosity defined in Eq.~(\ref{eqn:specLumi}). Since this specific luminosity is determined by the beam parameters, each detector and algorithm should measure identical values from the scan curve fits. 

For simplicity, the visible cross-section value extracted from a set of  {\em vdM} scans for a given detector and algorithm uses the convolved beam sizes measured by that same detector and algorithm.\footnote{An exception is the BCM\_EventAND algorithms, for which the visible cross-section is computed using the convolved beam sizes measured by the corresponding, higher-rate BCM\_EventOR algorithm, thereby providing slightly improved statistical accuracy.}
As shown in Fig.~\ref{fig:LSpec}, the values measured by LUCID\_EventOR and BCM\_EventOR are rather consistent within statistical uncertainties, 
although averaged over all BCIDs there may be a slight systematic difference between the two results.  
The difference observed between these two algorithms, after averaging over all BCIDs, results in a systematic uncertainty of 0.29\% related to
the choice of specific luminosity value.

\subsubsection{Length-scale calibration}

The length scale of each scan step enters into the extraction of $\Sigma_{x/y}$ and hence
directly affects the predicted peak luminosity during a {\em vdM} scan.
The length scale calibration procedure is described in Sect.~\ref{sec:LengthScale} and results 
in a $\pm 0.3\%$ uncertainty for scans IV--VIII.  For scans I--III, a less sophisticated length scale calibration 
procedure was performed which was more sensitive to hysteresis effects and re-centring errors resulting 
in a correspondingly larger systematic uncertainty of 2\%.

\begin{table*}
   \centering
   \caption{Relative systematic uncertainties on the determination of the visible cross-section $\sigma_\mathrm{vis}$ from {\em vdM} scans in 2010.  The assumed correlations of these parameters between scans is also indicated.}
   \label{tab:syst2010}
   \begin{tabular*}{\textwidth}{@{\extracolsep{\fill}}lcccl@{}} 
      	\hline
      	Scan Number & I & II--III & IV--V \\
      	Fill Number & 1059 & 1089 & 1386 \\
      	\hline
		Beam centring			& 2\% & 2\% & 0.04\% & Uncorrelated \\ 
		Beam-position jitter 		& -- & -- & 0.3\% & Uncorrelated \\
		Emittance growth\\
		\hspace{1em} and other non-reproducibility 	& 3\% & 3\% & 0.5\%  & Uncorrelated \\
		Fit model				& 1\% & 1\% & 0.2\% & Partially Correlated \\
		Length scale calibration	& 2\% & 2\% & 0.3\% & Partially Correlated \\
		Absolute length scale	& 0.3\% & 0.3\% & 0.3\% & Correlated \\
		Beam--beam effects       & -- & --  & 0.7\% & Uncorrelated \\
		Transverse correlations	& 3\%  &  2\%  & 0.9\% & Partially Correlated\\
		$\mu$ dependence		& 2\% & 2\% & 0.5\% & Correlated \\
		\hline
		Scan subtotal                      & 5.6\% & 5.1\% & 1.5\%  \\
		Bunch population product	& 5.6\% & 4.4\% & 3.1\% & Partially Correlated \\
		\hline
		Total					& 7.8\% & 6.8\% & 3.4\% \\
		\hline    
   \end{tabular*}
\end{table*}

\subsubsection{Absolute length scale of the Inner Detector}

The determination of the length scale relies on comparing the scan step requested by the LHC with
the actual transverse displacement of the luminous centroid measured by ATLAS.
This measurement relies on the length scale of the Inner Detector tracking system 
(primarily the pixel detector) being correct in measuring displacements of vertex positions
away from the centre of the detector.
An uncertainty on this absolute length scale was evaluated by analysing Monte Carlo events 
simulated using several different misaligned Inner Detector geometries.  
These geometries represent distortions of the pixel detector which are at the extreme limits of  
those allowed by the data-driven alignment procedure.
Samples were produced with displaced interaction points to simulate the transverse beam 
displacements  seen in a {\it vdM} scan.
The variations between the true and reconstructed vertex positions in these samples 
give a conservative upper bound of $\pm0.3\%$ on the uncertainty on the determination 
of $\sigma_{\mathrm{vis}}$ due to the absolute length scale.

\subsubsection{Beam--beam effects}

For given values of the bunch intensity and transverse convolved beam sizes, which are precisely measured, the deflection-induced orbit distortion and the relative variation of $\beta^*$ are both proportional to $\beta^*$ itself; they also depend on the fractional tune. Assigning a $\pm 20$\% uncertainty on each $\beta$-function value at the IP and a $\pm 0.02$ upper limit on each tune variation results in a $\pm 0.5$\% ($\pm 0.7$\%) uncertainty on $\sigma_{\mathrm{vis}}$ for 2011 (2010). This uncertainty is computed under the conservative assumption that $\beta$-function and tune uncertainties are correlated between the horizontal and vertical planes, but uncorrelated between the two LHC rings; it also includes a contribution that accounts for small differences between the analytical and simulated beam--beam-induced orbit distortions.

\subsubsection{Transverse correlations}
\label{sec:correlations}

The {\em vdM} formalism outlined in Sect.~\ref{subsec:vdMform} explicitly assumes that 
the particle densities in each bunch can be factorized into independent horizontal and vertical 
components such that the term $1/(2 \pi \Sigma_x \Sigma_y)$ in Eq.~(\ref{eqn:lumifin})
fully describes the overlap integral of the two beams.
If the factorization assumption is violated, the convolved beam width $\Sigma$ in one plane is no 
longer independent of the beam separation $\delta$ in the other plane, although a straightforward 
generalization of the {\em vdM} formalism still correctly handles an arbitrary two-dimensional 
luminosity distribution as a function of the transverse beam separation $(\delta_x, \delta_y)$, 
provided this distribution is known with sufficient accuracy.

Linear $x$-$y$ correlations do not invalidate the factorization assumption, but they can rotate the 
ellipse which describes the luminosity distribution away from the $x$-$y$ scanning planes 
such that the measured $\Sigma_x$ and $\Sigma_y$ values no longer accurately reflect 
the true convolved beam widths~\cite{bib:Cai}.
The observed transverse displacements of the luminous region during the scans 
from reconstructed event vertex data directly measure this effect, and a 0.1\% upper limit
on the associated systematic uncertainty is determined.
This uncertainty is comparable to the upper limit on the rotation of the luminous region 
derived during 2010 LHC operations from measurements of the LHC lattice functions by 
resonant excitation, combined with emittance ratios based on wire-scanner data~\cite{bib:emittanceSimon}.

More general, non-linear correlations violate the factorization assumption, and additional 
data are used to constrain any possible bias in the luminosity calibration from this effect.
These data include the event vertex distributions, where both the position and shape of
the three-dimensional luminous region are measured for each scan step, and the offset scan data from 
scan IX, where the convolved beam widths are measured with a fixed beam--beam 
offset of 160 $\mu$m in the non-scanning plane. 
Two different analyses are performed to determine a systematic uncertainty.

First, a simulation of the collision process, starting with single-beam profiles constructed from the sum of two three-dimensional Gaussian distributions with arbitrary widths and orientations, is performed 
by numerically evaluating the overlap integral of the bunches.
This simulation, which allows for a crossing angle in both planes, is performed for each scan step to predict the
geometry of the luminous region, along with the produced luminosity.
Since the position and shape of the luminous region during a beam-separation scan 
varies depending on the single-beam parameters~\cite{bib:Kozanecki:2008zz}, 
the simulation parameters are adjusted to provide a reasonable description of the mean and 
RMS width of the luminous region observed at each scan step in the May 2011 scans 
VII--IX (including the offset scan).
Luminosity profiles are then generated for simulated {\em vdM} scans using these 
tuned beam parameters, and analysed in the same fashion as the real {\em vdM} scan 
data, which assumes factorization.
The impact of a small non-factorization in the single-beam distributions is 
determined from the difference between the `true' luminosity from the simulated overlap integral at 
zero beam separation and the `measured' luminosity from the luminosity profile fits.
This difference is 0.1--0.2\% for the May 2011 scans, depending on the fitting model used. 
The number of events with vertex data recorded during the 2010 {\it vdM} scans is 
not sufficient to perform a similar analysis for those scans.

A second approach, which does not use the luminous region data, fits the observed luminosity 
distributions as a function of beam separation to a number of generalized, 
two-dimensional functions.  
These functions include non-factorizable functions constructed from multiple two-dimensional Gaussian 
distributions with possible rotations from the scan axes, and other functions where factorization 
between the scan axes is explicitly imposed.
By performing a combined fit to the luminosity data in the two scan planes of scan VII, 
plus the two scan planes in the offset scan IX, the relative difference between the non-factorizable
and factorizable functions is evaluated for 2011.
The resulting fractional difference on $\sigma_{\mathrm{vis}}$ is $0.5$\%.
For  2010, no offset scan data are available, but a similar analysis performed on scans IV and V 
found a difference of 0.9\%.

The systematic uncertainty associated with transverse correlations is taken as the largest effect 
among the two approaches described above, to give an uncertainty of $0.5$\% for 2011.  
For 2010, the $0.9$\% uncertainty is taken as the difference between 
non-factorizable and factorizable fit models.

\subsubsection{$\mu$ dependence}
\label{sec:mudep}

\begin{figure}
   \centering
   \includegraphics[width=\columnwidth]{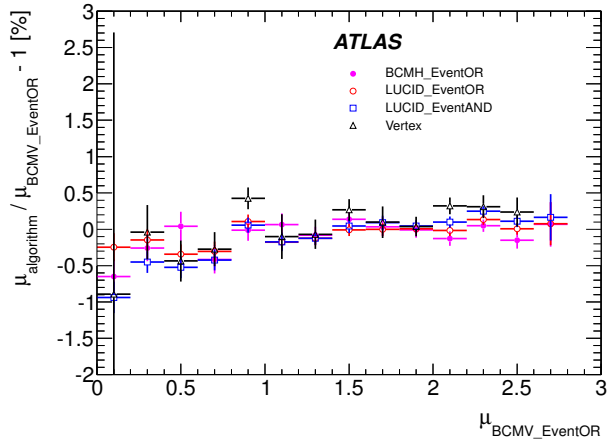} 
   \caption{Fractional deviation in the average value of $\mu$ obtained using different algorithms with respect to the BCMV\_EventOR value as a function of $\mu$ during scans VII--VIII.}
    \label{fig:mufig}
\end{figure}

Scans IV--V were taken over a range of interactions per bunch crossing $0 < \mu < 1.3$ while 
scans VII--VIII covered the range $0<\mu<2.6$, so uncertainties on the 
$\mu$ correction can directly affect the evaluation of $\sigma_{\mathrm{vis}}$.
Figure~\ref{fig:mufig} shows the variation in measured luminosity as a function 
of $\mu$ between several algorithms and detectors in 2011, and on the basis of this agreement
an uncertainty of $\pm 0.5\%$ has been applied for scans IV--VIII.\footnote{The number of interactions per bunch crossing $(\mu)$ is determined from the luminosity per bunch crossing as $\mu = {\cal L}\, \sigma_\mathrm{inel}/f_\mathrm{r}$ where the inelastic cross-section is assumed to be $\sigma_\mathrm{inel} = 71.5\, \mathrm{mb}$.  This value of $\sigma_\mathrm{inel}$ comes from a phenomenological model implemented in {\sc PYTHIA}6.4~\cite{Sjostrand:2006za} which was found to be consistent with the early luminosity calibrations in 2010~\cite{bib:ATLASLumiPaper}.  This cross-section is only used to present the luminosity data in terms of a more intuitive quantity, 
and does not enter into the luminosity determination in any way.}

Scans I--III were performed with $\mu \muchless 1$ and so uncertainties in the treatment of the 
$\mu$-dependent corrections are small.  
A $\pm 2\%$ uncertainty was assigned, however, on the basis of the agreement at low $\mu$ values between various detectors and algorithms, which were described in Ref.~\cite{bib:ATLASLumiPaper}.

\subsubsection{Bunch-population product}

The determination of this uncertainty has been described in Sect.~\ref{sec:currents}
and the contributions are summarized in Table~\ref{tab:BCNWG}.

\begin{table}[htbp]
   \centering
   \caption{Relative systematic uncertainties on the determination of the visible cross-section $\sigma_{\mathrm{vis}}$ from {\em vdM} scans in 2011.}
   \label{tab:syst2011}
   \begin{tabular*}{\columnwidth}{@{\extracolsep{\fill}}lc@{}}
      	\hline
      	Scan Number & VI--VII \\
      	Fill Number & 1783 \\
	\hline
		Beam centring		& 0.10\%  \\ 
		Beam-position jitter & 0.30\%  \\
		Emittance growth \\
		\hspace{1em} and other non-reproducibility		 	& 0.67\%   \\
		Bunch-to-bunch 
		$\sigma_{\mathrm{vis}}$ consistency & 0.55\% \\ 
		Fit model				& 0.28\%  \\ 
		Background subtraction   & 0.31\% \\
		Specific Luminosity           & 0.29\% \\
		Length scale calibration	& 0.30\%  \\
		Absolute length scale	& 0.30\% \\
		Beam--beam effects            & 0.50\% \\
		Transverse correlations	& 0.50\% \\
		$\mu$ dependence		& 0.50\% \\
		\hline
		Scan subtotal & 1.43\% \\
		Bunch population product	& 0.54\%  \\ 
		\hline
		Total					& 1.53\% \\
		\hline   
   \end{tabular*}
\end{table}

\subsection{Combination of 2010 scans}
\label{sec:combination}

The five {\em vdM} scans in 2010 were taken under very different conditions and have very different systematic uncertainties.
To combine the individual measurements of $\sigma_\mathrm{vis}$ from the five scans to determine the 
best calibrated $\overline{\sigma}_\mathrm{vis}$ value per algorithm, a Best Linear Unbiased Estimator 
(BLUE) technique has been employed taking into account both statistical and systematic 
uncertainties, and the appropriate correlations \cite{bib:Lyons, bib:BLUE}.
The BLUE technique is a generalization of a $\chi^2$ minimization, where for any set of 
measurements $x_i$ of a physical observable $\theta$, 
the best estimate of $\theta$ can be found by minimizing
\begin{equation}
\chi^2 = (\mathbf{x} - \bm{\theta})^\mathrm{T} \mathbf{V}^{-1} (\mathbf{x} - \bm{\theta})
\end{equation}
where $\mathbf{V}^{-1}$ is the inverse of the covariance matrix $\mathbf{V}$, and $\bm{\theta}$ is the product of the unit vector and $\theta$.

Using the systematic uncertainties described above, including the correlations indicated in Table~\ref{tab:syst2010}, a covariance matrix is constructed for each error source according to $V_{ij} = 
\sigma_i\, \sigma_j\, \rho_{ij}$ where $\sigma_i$ is the uncertainty from a given source for scan $i$, and $\rho_{ij}$ is the linear correlation coefficient for that error 
source between scans $i$ and $j$. As there are a total of five {\it vdM} scans, a $5\times 5$ covariance matrix is determined for each source of uncertainty.
These individual covariance matrices are combined to produce the complete covariance matrix, along with the statistical uncertainty shown in Table~\ref{tab:sigvis2010}.
While in principle, each algorithm and detector indicated in Table~\ref{tab:sigvis2010} could have different systematic uncertainties, no significant sources of systematic uncertainty have 
been identified which vary between algorithms.
As a result, a common systematic covariance matrix has been used in all combinations.

The best estimate of the visible cross-section $\overline{\sigma}_\mathrm{vis}$ for each luminosity method
in 2010 is shown in Table~\ref{tab:sigvisresult} along with the uncertainty. 
Because the same covariance matrix is used in all combinations aside from the small statistical component, 
the relative weighting of the five scan points is almost identical for all methods.
Here detailed results are given for the LUCID\_EventOR combination.
Because most of the uncorrelated uncertainties were significantly reduced from scans I--III to scans IV--V, the values from the last two {\it vdM} scans dominate the combination.
Scans IV and V contribute a weight of  45\% each, while the other three scans make up the remaining 10\% of the weighted average value.
The total uncertainty on the LUCID\_EventOR combination represents a relative error of $\pm3.4\%$, and is nearly identical to the uncertainty quoted 
for scans IV--V alone in Table~\ref{tab:syst2010}.
Applying the beam--beam corrections described in Sec.~\ref{subsec:beambeamCrctns}, which only affect scans IV--V 
in 2010, changes the best estimate of $\overline{\sigma}_\mathrm{vis}$ by $+1.9\%$ compared to making no corrections 
to the 2010 calibrations.

\begin{figure}
   \centering
   \includegraphics[width=\columnwidth]{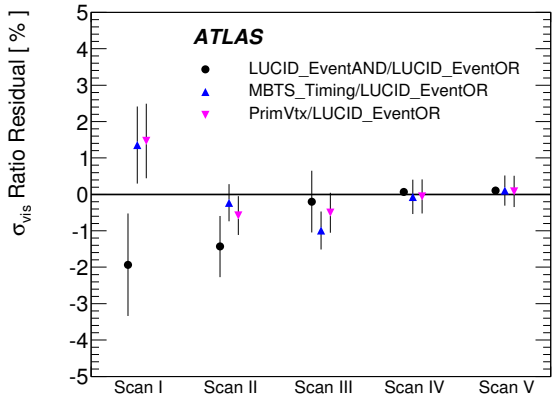} 
   \caption{Residuals of the $\sigma_\mathrm{vis}$ ratios between algorithms for each scan in 2010 are shown as a relative deviation from the mean ratio based on $\overline{\sigma}_\mathrm{vis}$.  Error bars represent statistical uncertainties only.}
   \label{fig:ratio}
\end{figure}

Figure~\ref{fig:ratio} shows the agreement among the algorithms within each scan 
in 2010 by plotting the deviations of the ratios $\sigma_\mathrm{vis}/\sigma_\mathrm{vis}(\mathrm{LUCID\_EventOR})$ for several algorithms from the mean value of these ratios, $\overline{\sigma}_\mathrm{vis}/\overline{\sigma}_\mathrm{vis}(\mathrm{LUCID\_EventOR})$.
By construction, any variation between scans related to the bunch population 
product $n_1 n_2$ cancels out, and the remaining scatter reflects the variation 
between algorithms in measuring $\mu_\mathrm{vis}^\mathrm{MAX} \Sigma_x \Sigma_y$.
The observed variation is mostly consistent with the statistical uncertainties, and the 
observed variation of up to $\pm2\%$ is consistent with the systematic uncertainty assigned 
to scans I--III for $\mu$ dependence.
No evidence for any additional source of significant systematic uncertainty between the 
algorithms is apparent.

\begin{table}
   \centering
   \caption{Best estimates of the visible cross-section determined from {\it vdM} scan data for 2010 and 2011.  Total uncertainties are shown including the statistical component and the total systematic uncertainty taking all correlations into account.  The 2010 and 2011 values are not expected to be consistent due to changes in the hardware for LUCID and BCM, and changes in the algorithm used for vertex counting.
}
   \label{tab:sigvisresult}   
   \begin{tabular*}{\columnwidth}{@{\extracolsep{\fill}}lr@{\extracolsep{0pt}$\,\pm\,$}l@{\extracolsep{\fill}}r@{\extracolsep{0pt}$\,\pm\,$}l@{}} 
      	\hline
      	 & \multicolumn{4}{c}{Visible cross-section $\overline{\sigma}_\mathrm{vis}$ (mb)}  \\
	 & \multicolumn{2}{c}{2010} & \multicolumn{2}{@{\extracolsep{\fill}}c}{2011} \\
      	\hline
		LUCID\_EventAND	& 13.3 &  0.5 & 13.7 &  0.2 \\  
		LUCID\_EventOR 	& 42.5 &  1.5 & 43.3 &  0.7  \\
		LUCID\_EventA         & \multicolumn{2}{c}{} & 28.5 &  0.4\\
		LUCID\_EventC        & \multicolumn{2}{c}{} & 28.5 &  0.4\\
		BCMH\_EventAND 	& 0.134 &  0.005 &  0.140 &  0.002 \\
		BCMV\_EventAND	& \multicolumn{2}{c}{} & 0.142 &  0.002  \\
		BCMH\_EventOR	& 4.69 &  0.16 & 4.78 &  0.07  \\ 
		BCMV\_EventOR	&  \multicolumn{2}{c}{} & 4.82 &  0.07  \\
		MBTS\_Timing		& 52.1 &  1.8 & \multicolumn{2}{c}{} \\
		PrimVtx			& 50.2 &  1.7 & \multicolumn{2}{c}{} \\
		Vertex (5 tracks)	& \multicolumn{2}{c}{} & 39.1 &  0.6  \\
		\hline
\end{tabular*}
\end{table}

\section{Luminosity extrapolation}
\label{sec:extrapolation}

The $\overline{\sigma}_{\mathrm{vis}}$ values determined in Sect.~\ref{sec:errors} allow each 
calibrated algorithm to provide luminosity measurements over the course of the 
2010 and 2011 runs.
Several additional effects due to the LHC operating with a large number of bunches 
and large $\mu$ values must be considered for the 2011 data, however, and additional 
uncertainties related to the extrapolation of the {\it vdM} scan calibration to the complete 
data sample must be evaluated.

Several specific corrections are described in this section for the 2011 data, while more general 
uncertainties, related to the agreement and stability of the various luminosity methods
applicable to both 2010 and 2011, are described in Sect.~\ref{sec:stability}.

\subsection{2011 hardware changes}
\label{sec:hardware}

Several changes were made to the readout chain of both the BCM and LUCID detectors 
before and during the early 2011 data-taking period. 

During the 2010--2011 LHC winter shutdown, resistors on the BCM front-end boards were replaced to
increase the dynamic range of the low-gain BCM signals used for beam-abort monitoring.
While the adjustments were performed in a way that should have left the high-gain BCM signal 
(used for the luminosity measurement) unchanged, variations at the percent level remain possible.
As a result, the BCM calibration in 2010 is not expected to be directly applicable to the 2011 data.

On 21 April 2011, the BCM thresholds were adjusted to place them at a better 
point in the detector response plateau.
As this change was made during a period with stable beams, the ratio of the BCM luminosity to that of any other detector
shows a clear step, which can be used to  measure directly the relative change in 
$\sigma_{\mathrm{vis}}$ due to this adjustment.
After the threshold change, the luminosity measured by BCMH\_EventOR was observed to 
increase with respect to other detectors by $+3.1\%$, which implies that the 
$\sigma_{\mathrm{vis}}$ value for BCMH\_EventOR decreased by this amount.  
For BCMV the equivalent luminosity change is $+4.1\%$.
Since the 2011 {\em vdM} scan calibration happened after this date, for any BCM data taken before 
this threshold change, the $\sigma_{\mathrm{vis}}$ values applied have been scaled up 
accordingly from the 2011 calibrated values.
The total change in  $\sigma_{\mathrm{vis}}$ for BCMH\_EventOR  shown in 
Table~\ref{tab:sigvisresult} is $+2.5\%$, implying that over the 2010--2011 winter shutdown 
the BCMH\_EventOR response changed by about $+5.6\%$. 

During the LHC technical stop in early April 2011, the LUCID receiver cards were changed 
to improve the performance of the readout with 50~ns bunch spacing.
Since this change was made during a period with no collisions, there is no direct measurement 
of the shift in LUCID calibration. 
Using data taken before and after the technical stop it can be estimated that the 
LUCID\_EventOR $\sigma_{\mathrm{vis}}$ value increased by about 2--3\%. 
The total change in LUCID\_EventOR calibration from 2010 to 2011 shown in 
Table~\ref{tab:sigvisresult} is $+2.4\%$, which indicates that the LUCID 
$\sigma_{\mathrm{vis}}$ calibration is consistent between 2010 and 2011 
at a level of approximately $1\%$.

Finally, on 30 July 2011, the radiator gas was removed from the LUCID Cherenkov tubes and 
the detector was operated for
the rest of the 2011 physics run using only the Cherenkov signal from the quartz window.
This reduction in detector efficiency was motivated by several factors, including the 
increasing interaction rate which was starting to saturate the 
LUCID\_EventOR response when the detector was filled with gas, 
as well as the better stability and linearity observed without gas.
The calibration of the LUCID luminosity measurements without gas was determined by comparing to the TileCal 
luminosity as described in Sect.~\ref{sec:lucidpmt}.

\subsection{Backgrounds}
\label{sec:backgrounds}

As described in Sect.~\ref{sec:vdmAnalysis}, 
both the LUCID and BCM detectors observe some small ``afterglow'' activity in 
the BCIDs immediately following a collision in normal physics operations.
With a 2011 bunch spacing of 50~ns and a relatively large number of bunches
injected into the LHC, this afterglow  tends to reach a fairly stable equilibrium after the first few 
bunches in a train, and is observed to scale with the instantaneous luminosity.

\begin{figure}[htbp] 
   \centering
   \includegraphics[width=\columnwidth]{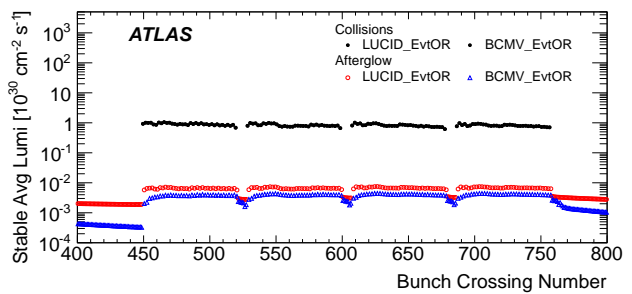} 
   \caption{Observed  luminosity  averaged over the fill as a function of BCID for the LUCID\_EventOR and 
   BCMV\_EventOR algorithms for a single LHC fill with 1042 colliding bunch pairs.
   On this scale the BCMV and LUCID luminosity values for colliding BCIDs are
    indistinguishable.
   The small ``afterglow'' luminosity comes in BCIDs where no bunches are colliding and is 
   the result of induced activity seen in the detectors.  Only 400 BCIDs are shown so that the 
   details of the afterglow in the short and long gaps in the fill pattern can be seen more clearly.}
   \label{fig:afterglow}
\end{figure}

Figure~\ref{fig:afterglow} shows the  luminosity as determined by \mbox{LUCID\_EventOR} 
and \mbox{BCMV\_EventOR} for a span of 400 BCIDs within a fill in June 2011 with 1042 colliding bunch pairs.
The afterglow level can be seen to be roughly constant at the $1\%$ level for 
 LUCID\_EventOR and at the $0.5\%$ level for BCMV\_EventOR during the bunch train, 
 and dropping during gaps in the fill pattern.
 
 To assess the effect of afterglow, the probability of an afterglow event must be combined with 
 the Poisson probabilities outlined in Sect.~\ref{sec:mu} to obtain the correction to the observed
 $\mu$ value.  
 For EventOR and HitOR algorithms, this correction is  
 $\mu = \mu_{\mathrm{obs}} - \mu_{\mathrm{bgd}}$ while for the EventAND algorithms a considerably 
 more involved formula must be applied.
 To estimate $\mu_{\mathrm{bgd}}$, the calibrated $\mu$ value observed in the BCID immediately 
 preceding a collision has been used.
 Different estimates using the following BCID or the average of the preceding and following BCIDs 
 produce negligibly different results.
 
 This afterglow subtraction has been applied to all BCM and LUCID luminosity determinations. 
 Since the afterglow level in the BCID immediately following a colliding bunch may be different 
 from the level in the second BCID after a colliding bunch, BCIDs at the end of a bunch train 
 have been used to evaluate any possible bias in the afterglow correction.
 It is observed that the simple afterglow subtraction over-corrects for the afterglow
 background in the BCMH\_EventOR algorithm by approximately 0.2\%, although for the 
 BCMV\_EventOR algorithm the method works better.
 A systematic uncertainty of $\pm 0.2\%$ is assigned to cover any possible bias on the BCMV\_EventOR luminosity.
The LUCID\_EventOR algorithm is over-corrected by around 0.5\%, and this
bias is removed by applying a constant scale factor to the LUCID luminosity measurements.
 A more detailed comparison, using luminosity data from a single-bunch run to
 construct an afterglow ``template'' which can be combined with any arbitrary 
 bunch pattern to emulate the behavior in a train, yields consistent results.

 Afterglow in 2010 was considerably less important due to the 150~ns bunch spacing, and the 
 relatively short trains used that year.
 Afterglow is generally negligible in {\em vdM} scans due to the small number of colliding bunches  
 and the large spacing between them.

The additional single-beam backgrounds observed by both BCM and LUCID are generally negligible
during normal physics operations as these luminosity-independent backgrounds are tiny compared to the
typical signal during physics operations.
These backgrounds must be treated carefully, however, during {\em vdM} scans
or other special beam tests which involve low-luminosity running.

\subsection{LUCID PMT current correction}
\label{sec:lucidpmt}

Due to the increase in the total luminosity delivered by the LHC, both in terms of the number 
of bunches colliding and of the average number of interactions per bunch crossing, the LUCID PMTs in 2011 were 
operating in a regime where the average anodic PMT current is of order $10\,\mu$A, which has an 
observable effect on the PMT gain.

Uncorrected, this effect shows up both as an apparent $\mu$ dependence of the 
luminosity, since the PMT currents are highly correlated with the average $\mu$ during a fill, 
as well as a long-term time dependence in the LUCID luminosity value, since the number
of colliding bunches steadily increased in 2011.
The magnitude of this effect was of the order of $4\%$ on the LUCID\_EventOR luminosity by the end of 2011.

The total anodic current summed over all LUCID tubes has been observed to produce a 
deviation of the luminosity measured by the various LUCID algorithms with respect to the TileCal value.
A correction for this effect has been evaluated using a single ATLAS run with 1317 colliding
bunches.
TileCal is used as the reference, and a second-order polynomial is fitted to the ratio between the 
LUCID and TileCal luminosity, for all the algorithms, as a function of the total anodic PMT current.
This PMT current correction has been applied to all LUCID data used to 
describe luminosity during physics operations.

The constant term of the fitted function, representing the extrapolation to zero PMT anodic current,
provides the correction to be applied to the LUCID {\em vdM} calibration resulting
from the removal of the radiator gas from the detector, as well as from any ageing-related variation 
in PMT gain to that point in time. 
As discussed in Sect.~\ref{sec:particlealg}, the TileCal luminosity calibration is performed relative to 
LUCID\_EventOR at the time of the {\em vdM} scan.
As a result, the LUCID and TileCal luminosity measurements are implicitly tied to each other at one point
in time, although any long-term variations away from that point are still significant.
Similarly, any $\mu$ dependence between the LUCID and TileCal response is 
largely removed by this correction procedure, although comparisons to other 
detectors remain relevant.

\subsection{BCM calibration shifts}
\label{sec:bcmdrift}

The BCM detectors are solid-state devices constructed from chemical vapour deposition 
diamonds to provide tolerance to high radiation levels.
A well-known feature of such detectors is a tendency for the gain to increase under 
moderate irradiation levels up to a stable
asymptotic value at high dose rates~\cite{Adam:2006kz, Adam:2000ra}.
This so-called ``pumping'' is generally ascribed to the filling of charge traps in the diamond 
sensors with continued irradiation until enough charge has been sent through the device to 
fill essentially all of the traps.
Measurements of this effect in diamond samples outside ATLAS and
the predicted fluences in the presence of LHC collisions predict that
the diamonds should become fully pumped within tens of minutes when the ATLAS
instantaneous luminosity is $10^{33}\, \mathrm{cm}^{-2}\, \mathrm{s}^{-1}$.

\begin{figure}
   \centering
   \includegraphics[width=\columnwidth]{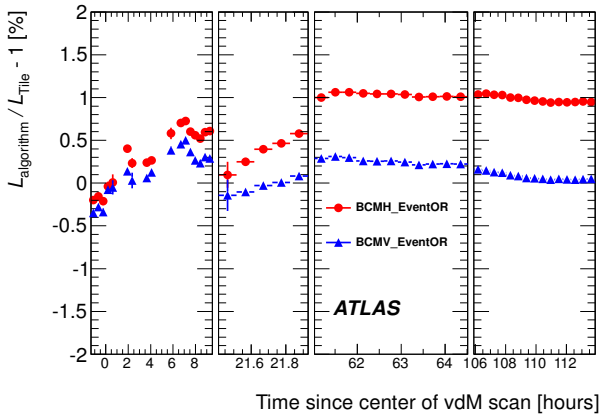} 
   \caption{Fractional deviation of BCMH\_EventOR and BCMV\_EventOR luminosity values with respect to TileCal as a function of time since the May 2011 {\em vdM} scan.  
   The TileCal luminosity scale is calibrated to LUCID\_EventOR at the time of the {\em vdM} scan.
   The {\em vdM} scan was performed immediately following an LHC technical stop, when there had
    been no collisions for about 2 weeks.
    }
   \label{fig:bcmdrift}
\end{figure}

In the 2011 BCM data it has been observed that the apparent luminosity scale of the 
different sides of the BCM detectors tends to vary by up to about 1\% 
immediately after an extended period with no beam in the LHC. 
Figure~\ref{fig:bcmdrift} shows the fractional deviation of the BCMH\_EventOR and 
BCMV\_EventOR luminosity values from the luminosity measured by TileCal.  
The {\em vdM} calibration occurs near the start of the period shown in this figure, and a clear 
drift of the BCMH\_EventOR luminosity scale is observed during the first fill and the start of the
second fill, until settling at an asymptotically stable value.
The drift of the BCMH\_EventOR luminosity from the calibrated value is estimated to be 
$+1.0\%$, while the BCMV\_EventOR luminosity is consistent with no significant net drift by 
the end of this time interval.
Comparable shifts are observed in the BCM\_EventAND luminosity scales. 
Similar patterns are observed after each LHC technical stop, 
a two or three week period during physics running, scheduled approximately every two months to allow for machine development and equipment maintenance.
Within a couple of fills after each technical stop has ended and normal physics collisions 
have resumed, the BCM luminosity scale is observed to return, with rather good reproducibility,
to the level recorded before the technical stop.

One interpretation of these data is that a small amount of annealing at the few percent level
can occur during the technical stops.
In the first few low-luminosity fills after a technical stop, some amount 
of ``micro-pumping'' takes place to refill these short-lifetime traps.
The first fill shown in Fig.~\ref{fig:bcmdrift} is the {\em vdM} scan, which takes place 
right  after the May 2011 technical stop. 
With an average luminosity around $3\times 10^{30}\, \mathrm{cm}^{-2}\, \mathrm{s}^{-1}$, this fill does not provide enough 
particle  fluence through the BCM detectors to fully pump the short-lifetime traps.
By the time of the third fill, where the luminosity reaches  $4\times 10^{32}\, \mathrm{cm}^{-2}\, \mathrm{s}^{-1}$, the particle fluences since the  technical stop are sufficient to return the detectors to their asymptotic response.

To account for this short-term change in the BCMH detector response, the BCMH luminosity scale
has been corrected by the observed 1.0\% drift after the {\em vdM} scan.
No correction has been applied to the BCMV\_EventOR algorithm which is used to set the
physics luminosity scale, but an additional systematic uncertainty of $\pm 0.25\%$ has been 
applied as an estimate of the uncertainty due to this effect.

\subsection{TileCal calibration}
\label{sec:tilecalibration}

As described in Sect.~\ref{sec:particlealg}, the TileCal PMT currents from selected cells are calibrated
with respect to the luminosity observed by the LUCID\_EventOR algorithm at relatively low $\mu$ values.
This current-based luminosity measurement is not absolutely calibrated, and does not provide 
bunch-by-bunch information, but is still a valuable cross-check of the stability of the other 
luminosity algorithms. 

In the 2010 data, the total TileCal PMT current  for a common group of cells is calibrated during a 
single LHC fill taken in October 2010.
The calibration is performed by fitting the TileCal response as a function of the LUCID\_EventOR luminosity
over a range \mbox{$50$--$100\times10^{30}~\mathrm{cm}^{-2}~\mathrm{s}^{-1}$} with a first-order polynomial, where the
constant term accounts for any pedestal or non-collision backgrounds present in the TileCal currents.
This cross-calibrated luminosity value is then compared to  LUCID\_EventOR for all of the 
2010 $pp$ data where the luminosity was greater than 
$35\times10^{30}~\mathrm{cm}^{-2}~\mathrm{s}^{-1}$. 
This luminosity represents the approximate threshold above which the luminosity-based current
signal is large enough to be resolved.
The RMS residual deviation between TileCal and LUCID is found to be about 
0.2\% when comparing the average luminosity measured over a time range of about 2 minutes.

For the calibration method used in the 2011 data, a few cells around $|\eta| = 1.25$ with the highest observed 
currents are compared to the LUCID\_EventOR luminosity at the peak of the {\em vdM} scan.
The TileCal pedestals are explicitly measured using data taken at the start of the fill before the beams are
put into collision, and the pedestal-corrected TileCal currents are assumed to be directly proportional to the luminosity (with no constant offset).
The LUCID luminosity at the peak of the {\em vdM} scan, which is itself calibrated by the scan at that point in time, is simply used
to set the proportionality constant for each TileCal cell.
These few calibrated cells are then compared to other TileCal cells in a fill shortly after the {\em vdM} scan when the luminosity
is in the range \mbox{$100$--$200\times10^{30}~\mathrm{cm}^{-2}~\mathrm{s}^{-1}$} which is high enough to produce a reasonable current in all cells.  
The proportionality constants for these remaining cells are determined by comparing the pedestal-corrected currents to the luminosity measured by the subset of cells which were directly calibrated during the {\em vdM} scan. 
This two-stage calibration is necessary because the total luminosity during the {\em vdM} scan is too low to provide 
reasonable currents to all of the TileCal cells used to measure luminosity.
The result is a TileCal calibration which is nearly 
independent of LUCID or any other detector in 2011.

The calibration of individual cells in 2011 allows all available cells to be used at any given time to provide a luminosity,
 which is important in 2011 due to an increasing number of tripped TileCal cells over the course of the year.
 Since the set of available cells can vary significantly over time, this method is more sensitive to 
the residual variations of the cell calibration constants. 
For the 2011 data, the RMS variation of the TileCal luminosity measurement is estimated to 
be about $0.5\%$ based on the agreement between individual cells and the typical number of 
calibrated cells available to make a measurement.

Additionally, the response of the TileCal PMTs showed variations in time related to the exposure of the detector to collisions.
A downward drift of the mean PMT response was observed during data-taking periods, and an upward drift back to an asymptotically stable
value was observed after a few days during a technical stop when there were no collisions.
The typical size of this variation is around $1\%$.
This effect has been identified during calibration runs with a caesium-137 source that circulates among 
the TileCal cells and during laser calibration runs, where a laser signal is directly injected into 
the PMTs.
Comparison of the luminosity measured by specific TileCal cells also confirms a time variation based on the rates of exposure seen by each 
individual cell.
The TileCal laser calibration system is used to derive a global correction factor as a function of time based on the observed change 
in mean PMT response.
This global correction improves the time stability of the TileCal luminosity, but as discussed further in Sect.~\ref{sec:timestability} it does
not remove the effect completely.
Performing cell-by-cell corrections is unfeasible as the statistical error on the individual cell corrections would be too large.

\subsection{FCal calibration}
\label{sec:fcalcalibration}

Similarly, the FCal high-voltage (HV) currents are calibrated to one of the other detectors 
at one time to provide a  luminosity measurement which can be used to  check the 
stability of other methods. The FCal needs a higher instantaneous luminosity than TileCal 
(a minimum value around $1 \times10^{32}~\mathrm{cm}^{-2}~\mathrm{s}^{-1}$) to have a significant current signal.
In order to check the validity of the calibration throughout the 2010 data-taking period, the 
calibrated FCal luminosity is compared to the LUCID\_EventOR luminosity for a set of runs recorded during 
October 2010 when the luminosity was high enough for the FCal technique to work.
The RMS residual variation between FCal and LUCID is found to be about 0.5\%.
For 2011, a similar calibration was performed between FCal and BCMV\_EventOR during a single run.
The FCal HV lines are selected for luminosity determination based on their noise, and
lines that are connected to shorted calorimeter electrodes are excluded.
Individual HV currents are then compared to BCMV\_EventOR during an LHC fill in September 
when the beams were purposely separated to provide a wide range of $\mu$ values in a short
period of time.
These so-called ``$\mu$ scans'' are also used to assess the $\mu$ dependence of various
algorithms as described in Sect.~\ref{sec:mudependence}.
The $\mu$-scan data provide the largest range of luminosities to calibrate the FCal 
current data accurately, and a linear fit is applied to extract calibration 
parameters for each FCal HV line.
These calibrations are then applied to all measured HV currents in 2011 to provide a measured 
luminosity per HV line, and these individual measurements are averaged to produce a single FCal 
luminosity measurement.

\section{Luminosity stability}
\label{sec:stability}

To produce the integrated luminosity values used in ATLAS physics analyses, a single algorithm is 
chosen to provide the central value for a certain range of time, with the remaining calibrated 
algorithms providing independent measurements to evaluate systematic uncertainties 
on the stability of these results.
The LUCID\_EventOR algorithm is primarily used in 2010 where the large visible cross-section 
makes it more sensitive to the relatively low luminosity delivered in that year. 
In 2011 the  BCMV\_EventOR algorithm is primarily used, due to the better relative stability of this 
detector compared to either BCMH or LUCID during the 2011 run.

The calibration of $\sigma_{\mathrm{vis}}$ is performed on only a few occasions 
(only once in 2011) and at a relatively low value of $\mu$ compared to the range of 
$\mu$ values routinely seen in physics operations, particularly in 2011 where peak values
of $\mu\simeq 20$ for certain BCIDs were not uncommon.
As discussed in Sec.~\ref{sec:mudep}, the number of interactions per bunch crossing $(\mu)$ is 
equivalent to the luminosity per bunch crossing and provides an intuitive unit to describe pile-up conditions.

Two additional sources of uncertainty are evaluated, which are related to the stability of the calibrated
results when applied to the entire 2010 and 2011 data samples.
The first is the long-term  stability of each algorithm with respect to time, and the second is the
linearity of the calibrated luminosity value with respect to the interaction rate $\mu$.
In each case, the agreement between all available detectors and algorithms is used to limit 
the possible systematic variation of the primary algorithm used to deliver physics luminosity results.

\subsection{Long-term stability}
\label{sec:timestability}

One key source of potential uncertainty is the assumption that the 
$\overline{\sigma}_{\mathrm{vis}}$ 
calibration determined in a set of {\em vdM} scans is stable across the entire 2010 or 2011 
data set.
Several effects could degrade the long-term stability of a given detector, including slow drifts
in the detector response and sensitivity to varying LHC beam conditions, particularly the 
total number of colliding bunches.
Because the number of colliding bunches increased rather monotonically during both the 
2010 and 2011 data-taking periods, it is not  possible to disentangle these two effects, 
so the tests of long-term stability should be viewed as covering both possibilities.

\begin{figure}
   \centering
  \includegraphics[width=\columnwidth]{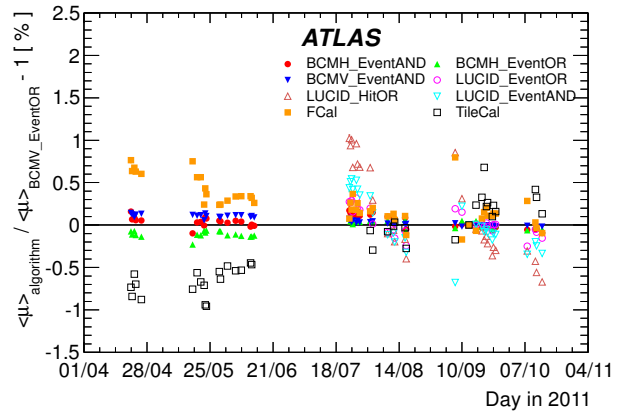} 
   \caption{Fractional deviation of the mean interaction rate obtained using different algorithms from the BCMV\_EventOR value as a function of time in 2011.  
   Each point shows the mean deviation of the rate in a single run from the rate 
   in a reference run 
   taken in the middle of September.  
   Statistical uncertainties per point are negligible.}
   \label{fig:muvstime}
\end{figure}

Figure~\ref{fig:muvstime} shows the interaction rate ratio of a given algorithm to the 
reference algorithm as a function of time in 2011. 
Each point shows the average number of interactions per bunch crossing measured by a particular algorithm divided by the number measured by  BCMV\_EventOR, averaged over one ATLAS run.
The average number of interactions per bunch crossing, $\langle\mu\rangle$, is the number of interactions per bunch $\mu$ averaged over all BCIDs
with colliding bunch pairs, and must be used for any comparison with TileCal or FCal.
The figure shows the relative variation of this ratio over time compared to a single fill in 
September which is used to provide a reference point, and comes approximately four months after the {\em vdM} scan in May.
The variation seen on the left-hand side of this plot indicates the level of long-term stability
from the {\em vdM} scan until this time in mid-September.

The various BCM algorithms are very stable with respect to each other, with agreement at the 
level of a few tenths of a percent over the entire 2011 run
(the first few fills with low numbers of colliding bunches after each technical stop are not shown 
in this figure).
This demonstrates the reproducibility of the BCM luminosity scale after
each technical stop as discussed in Sect.~\ref{sec:bcmdrift}.
The LUCID data are shown only for the period of operation without gas from July onwards.
Some variation at the level of $\pm 0.5\%$ can be seen for the LUCID\_Event algorithms, 
with somewhat larger variations observed for LUCID\_HitOR.
These variations are observed to be correlated with drifts in the PMT gains inferred from measurements 
of single-photon pulse-height distributions in the LUCID data.

The FCal luminosity scale is observed to change by about $-0.5\%$ with respect to BCMV\_EventOR
from early to late 2011.
Studies have shown that this variation is actually the result of a residual non-linearity in the FCal 
luminosity response.
Since the average luminosity increased considerably from early to late 2011 due to the 
increase in the number of colliding bunches, this non-linearity with total luminosity 
manifests itself as an apparent drift on the time stability plot.
The TileCal luminosity is observed to undergo a slow drift with respect to BCMV\_EventOR at
the level of 1\% over the course of 2011. 
In contrast to the FCal, this variation has been shown not to be dependent on luminosity,
but rather is likely due to residual PMT gain variations which are not corrected by the TileCal 
laser calibration system.

Based on the observed variation with time between the various algorithms shown in 
Fig.~\ref{fig:muvstime}, a systematic uncertainty on long-term stability, which includes 
any effects related to dependence on the number of colliding bunches or other 
operational conditions seen in the 2011 data, is set at $\pm 0.7\%$.
Similar tests on the 2010 data show consistency at the level of $\pm 0.5\%$, 
where very good agreement is observed between the LUCID, BCM, TileCal, and FCal luminosity
measurements.

\subsection{Interaction rate dependence}
\label{sec:mudependence}

A final key cross-check is the level of agreement between the calibrated luminosity algorithms as a function of $\mu$, 
the number of interactions per bunch crossing.
In 2010, the measured values of $\mu$ in normal physics operations were in the range $0 < \mu < 5$, and
a direct comparison of the four LUCID and BCMH algorithms over this range showed agreement
at the $\pm 0.5\%$ level.
In 2011, the measured values of $\mu$ seen in physics data are considerably larger, with most data in the range
$4 < \mu < 20$.
The effects of pile-up increase at larger interaction rates, and it is important to verify that the various algorithms still provide an accurate and
linear measurement of the luminosity up to the highest values of $\mu$ observed in the data.

\begin{figure}
   \centering
   \includegraphics[width=\columnwidth]{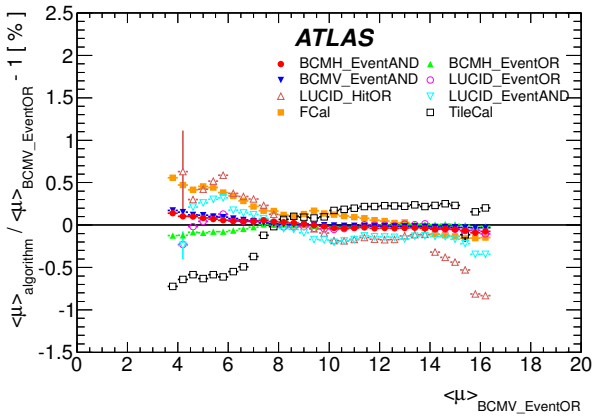} 
   \caption{Fractional deviation of the average number of interactions per bunch crossing $\langle\mu\rangle$ (averaged over BCIDs) obtained using different algorithms from the BCMV\_EventOR value as a function of $\langle\mu\rangle$.  
Statistical uncertainties are shown per point, but generally are negligible.
 }
   \label{fig:muPlot}
\end{figure}

A first way to assess the linearity is to take the data presented in 
Fig.~\ref{fig:muvstime} and calculate the  interaction rate ratio as a function of the average 
number of interactions per bunch crossing $\langle\mu\rangle$.
This is shown in Fig.~\ref{fig:muPlot}.
Because the calorimeter methods measure only the interaction rate averaged over
all colliding bunches $\langle\mu\rangle$, the range of this comparison is smaller than
the BCID-sensitive methods which test the full $\mu$ range.
Since there is no absolute linearity reference available, the agreement between multiple 
algorithms with different acceptances and analysis methods is used to demonstrate 
consistency with each other, under the assumption that it is highly unlikely that they would
all deviate from linearity in exactly the same way.

Again, since there is a ramp-up in the number of interactions per bunch crossing with time in 2011, issues with time stability are reflected in this figure as an 
apparent $\langle\mu\rangle$ dependence.
The large variation in TileCal is a good example, as the data with 
$\langle\mu\rangle < 8$ were recorded largely before the July technical stop, while the data with $\langle\mu\rangle>8$ came mostly after this technical stop.
The FCal variation appears to be a genuine non-linearity, although studies show that this is
most accurately described as a dependence on total luminosity (not $\langle\mu\rangle$).
The LUCID\_HitOR response varies by up to $\pm 0.5\%$, although this is also most likely 
explained by the variations seen in the time stability.
The remaining algorithms all agree at the level of $\pm 0.5\%$, although this distribution does not
test the linearity of the algorithms all the way down to the {\em vdM} calibration at $\mu \approx 2$.

To improve the characterization of the $\mu$ dependence in the range $2 < \mu < 10$,
without complications from long-term stability,
a series of ``$\mu$-scans'' was performed in 2011 to provide a direct measurement of the
linearity of the various luminosity algorithms.
The $\mu$-scans are performed at the end of normal physics operations by separating the
beams by $\pm 5~\sigma_\mathrm{b}$ in $19$ steps, using the same procedure as employed in the
{\em vdM} scans.
Because this was done at the end of an LHC fill when the luminosity is fairly modest, and the
entire scan can be performed in less than an hour, the cost of this procedure in terms of lost 
physics luminosity is much less than performing a {\em vdM} scan.

During these $\mu$-scans, special triggers are used to collect large samples of
events for the vertex-based luminosity algorithm from two specific BCIDs.
In addition to the online algorithms, the TileCal and FCal current measurements also 
provide useful data during these scans.

\begin{figure}
   \centering
      \includegraphics[width=\columnwidth]{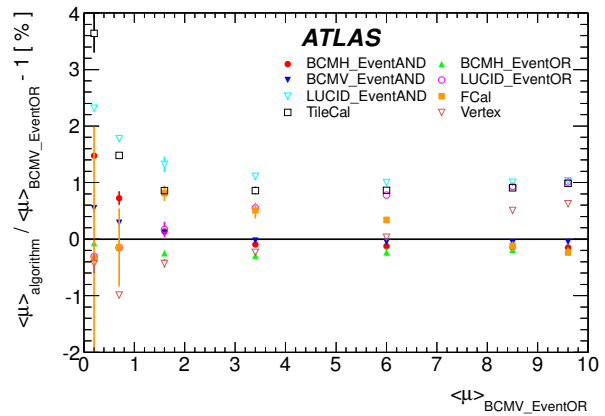} 
   \caption{Fractional deviation of the average number of interactions per bunch crossing $\langle\mu\rangle$ (averaged over BCIDs) obtained using different algorithms from the BCMV\_EventOR value as a function of $\langle\mu\rangle$.  Data shown are taken during a $\mu$-scan, where the beams are purposely separated to sample a large $\mu$ range under similar conditions.
     Statistical uncertainties are shown per point, but generally are negligible for $\langle\mu\rangle > 2$. 
}
   \label{fig:pileupScan}
\end{figure}

Figure~\ref{fig:pileupScan} shows the $\mu$-scan data comparison for several algorithms.  
Because single-beam backgrounds become relatively more important as the beams are separated, 
the LUCID and BCM data were corrected for both afterglow and single-beam backgrounds using a 
procedure similar to that employed in the {\em vdM} scans.

The approximately constant offsets between algorithms are the result of drifts in the calibrated scales due to long-term stability.
The linearity consistency is assessed by looking for a slope in the luminosity ratio with respect to the reference algorithm BCMV\_EventOR.
All of the algorithms show good linearity from the $\langle\mu\rangle$ value where the {\em vdM} scan is performed (around $\langle\mu\rangle=2$) up to the 
$\langle\mu\rangle$ value observed in nominal physics operations (here around 
$\langle\mu\rangle = 10$).
A deviation of around $1\%$ is observed in the FCal luminosity over this range,
which is consistent with the dependence on total luminosity also observed in Fig.~\ref{fig:muPlot}.
The TileCal data agree very well with BCM, which is significant since the TileCal luminosity scale
is cross-calibrated to LUCID\_EventOR during the {\em vdM} scan taken four months earlier.
The LUCID\_EventOR data also agree with BCM at the $\pm 0.5\%$ level, while 
LUCID\_EventAND deviates by a few percent at the lowest luminosity values.
This is interpreted as an imperfect subtraction of the single-beam background which is 
complicated by the presence of afterglow in this physics-based LHC filling pattern.
Deviations of LUCID\_EventAND are not observed at low luminosity in the {\em vdM} scan, 
shown in Fig.~\ref{fig:mufig}, where the background correction can be performed more
accurately.
The vertex counting data are also shown in Fig.~\ref{fig:pileupScan} for the two BCIDs which were 
recorded with a special trigger during this time.
The vertex luminosity increases by about 1\% over the range of this figure, which is consistent with 
the additional systematic uncertainties on the vertex counting technique.  These uncertainties, 
related to the vertex masking and fake vertex corrections, grow with the interaction rate and are estimated to reach $\pm 2\%$ by an interaction rate of $\mu =10$.

\begin{figure}
   \centering
      \includegraphics[width=\columnwidth]{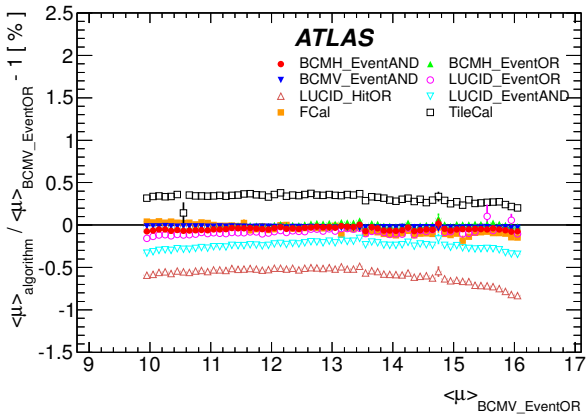} 
   \caption{Fractional deviation of the average number of interactions per bunch crossing $\langle\mu\rangle$ (averaged over BCIDs) obtained using different algorithms from the BCMV\_EventOR value as a function of $\langle\mu\rangle$.  Data from only a single LHC fill are shown.  
 Statistical uncertainties are shown per point, but generally are negligible.
 }
   \label{fig:singleFill}
\end{figure}

A final test of $\mu$ dependence is performed by comparing the luminosity ratio between
algorithms as a function of $\langle\mu\rangle$ for a single LHC fill.
This comparison, shown in Fig.~\ref{fig:singleFill}  for a fill in October 2011, provides a way to assess the linearity independently from any long-term stability effects up to the very highest $\mu$ 
values observed in 2011.
Here the shapes of the curves are directly sensitive to variations in the linearity as a function of 
$\langle\mu\rangle$, while the overall shifts of each algorithm up or down result from variations in the long-term stability.
So while TileCal and LUCID\_HitOR luminosity scales are both seen to deviate from 
BCMV\_EventOR by up to $0.5\%$, this variation is expected from the data shown in 
Fig.~\ref{fig:muvstime}.
Each algorithm shows a linear response with respect to BCMV\_EventOR, with the largest 
variations observed for LUCID\_HitOR at the $0.5\%$ level.

As a result of all the information available, a systematic uncertainty of $\pm 0.5\%$ has 
been applied to account for any possible $\mu$ dependence in the extrapolation from 
the low-$\mu$ {\em vdM} scan calibration to the higher-$\mu$ physics data in 2011.
More limited data were available in 2010, although the extrapolation range was 
significantly smaller ($\mu \le 5$).
Similar comparisons for the 2010 data lead to an uncertainty due to a possible $\mu$ dependence of $\pm 0.5\%$.

\subsection{Total systematic uncertainty}

Table~\ref{tab:errors} lists the contributions to the total systematic uncertainty on the luminosity 
scale provided for physics analyses in the 2010 and 2011 data samples.
The bunch population product and other calibration uncertainties are related to the {\em vdM} 
scan calibration described in Sects.~\ref{sec:calibration} and \ref{sec:errors}.
The afterglow and BCM stability uncertainties are related to particular conditions in 2011 as described in Sect.~\ref{sec:extrapolation}.
The long-term stability and $\mu$ dependence uncertainties are both related to extrapolating the {\em vdM} calibration to the entire 2010 and 2011 data samples as described in 
Sect.~\ref{sec:stability}.
The single largest improvement between 2010 and 2011 has come from a better understanding
of the bunch population product during the {\em vdM} scan.

\begin{table}
   \centering
   \caption{Relative uncertainty on the calibrated luminosity scale broken down by source.
   The {\it vdM} scan calibration uncertainty has been separated into the uncertainty on the
   bunch population product and the uncertainties from all other sources.}
   \label{tab:errors}
   \begin{tabular*}{\columnwidth}{@{\extracolsep{\fill}}lcc@{}}
      	\hline
	Uncertainty Source &  \multicolumn{2}{c}{$\delta {\cal L}/{\cal L}$} \\
	& 2010 & 2011 \\
	\hline
	         Bunch Population Product               & 3.1\% & 0.5\% \\
		Other {\it vdM} \\
		\hspace{1em}Calibration Uncertainties & 1.5\% & 1.4\% \\ 
		Afterglow Correction                   & & 0.2\% \\
		BCM Stability                                &  & 0.2\% \\
		Long-Term Stability			& 0.5\% & 0.7\% \\
		$\mu$ Dependence			& 0.5\% & 0.5\% \\
		\hline
		Total						& 3.5\% & 1.8\% \\
		\hline   
   \end{tabular*}
\end{table}

\section{Conclusions}
\label{sec:conclusions}
The luminosity scales determined by the ATLAS Collaboration for 2010 and 2011 have 
been calibrated based on data from dedicated beam-separation scans, also known as  van der Meer ({\it vdM}) scans.
Systematic uncertainties on the absolute luminosity calibration have been evaluated. 
For the 2010 calibrations,
the uncertainty is dominated by the understanding of the bunch charge product, while for 2011 
the uncertainty is mostly due to the accuracy of the {\em vdM} calibration procedure.
Additional uncertainties are evaluated to assess the stability of the calibrated luminosity scale over time
and over variation in operating conditions, most notably the number of interactions per bunch crossing.
The combination of these systematic uncertainties results in a final uncertainty on the 
ATLAS luminosity scale during $pp$ collisions at $\sqrt{s} = 7$~\TeV\  of 
$\delta {\cal L}/ {\cal L} = \pm 3.5\%$ for the $47\, \mathrm{pb}^{-1}$ of data 
delivered to ATLAS in 2010 and $\delta {\cal L}/ {\cal L} = \pm 1.8\%$ for the $5.5\, \mathrm{fb}^{-1}$ delivered in 2011.
These results include explicit corrections for beam--beam effects in the {\em vdM} calibration scans
that were not understood until late in the luminosity analysis and were therefore not applied to the luminosity scale used
in any ATLAS publication prior to July of 2013.
Consequently, the luminosity scale used in previous ATLAS results should
be scaled down by $1.9\%$ in 2010 and $1.4\%$ in 2011.
%

\section{Acknowledgements}

We thank CERN for the very successful operation of the LHC, as well as the
support staff from our institutions without whom ATLAS could not be
operated efficiently.

We acknowledge the support of ANPCyT, Argentina; YerPhI, Armenia; ARC,
Australia; BMWF and FWF, Austria; ANAS, Azerbaijan; SSTC, Belarus; CNPq and FAPESP,
Brazil; NSERC, NRC and CFI, Canada; CERN; CONICYT, Chile; CAS, MOST and NSFC,
China; COLCIENCIAS, Colombia; MSMT CR, MPO CR and VSC CR, Czech Republic;
DNRF, DNSRC and Lundbeck Foundation, Denmark; EPLANET, ERC and NSRF, European Union;
IN2P3-CNRS, CEA-DSM/IRFU, France; GNSF, Georgia; BMBF, DFG, HGF, MPG and AvH
Foundation, Germany; GSRT and NSRF, Greece; ISF, MINERVA, GIF, DIP and Benoziyo Center,
Israel; INFN, Italy; MEXT and JSPS, Japan; CNRST, Morocco; FOM and NWO,
Netherlands; BRF and RCN, Norway; MNiSW, Poland; GRICES and FCT, Portugal; MERYS
(MECTS), Romania; MES of Russia and ROSATOM, Russian Federation; JINR; MSTD,
Serbia; MSSR, Slovakia; ARRS and MVZT, Slovenia; DST/NRF, South Africa;
MICINN, Spain; SRC and Wallenberg Foundation, Sweden; SER, SNSF and Cantons of
Bern and Geneva, Switzerland; NSC, Taiwan; TAEK, Turkey; STFC, the Royal
Society and Leverhulme Trust, United Kingdom; DOE and NSF, United States of
America.

The crucial computing support from all WLCG partners is acknowledged
gratefully, in particular from CERN and the ATLAS Tier-1 facilities at
TRIUMF (Canada), NDGF (Denmark, Norway, Sweden), CC-IN2P3 (France),
KIT/GridKA (Germany), INFN-CNAF (Italy), NL-T1 (Netherlands), PIC (Spain),
ASGC (Taiwan), RAL (UK) and BNL (USA) and in the Tier-2 facilities
worldwide.

%
\bibliographystyle{spphys}
\raggedright
\bibliography{lumipaper}
\onecolumn 
\clearpage 
\begin{flushleft}
{\Large The ATLAS Collaboration}

\bigskip

G.~Aad$^{\rm 48}$,
T.~Abajyan$^{\rm 21}$,
B.~Abbott$^{\rm 111}$,
J.~Abdallah$^{\rm 12}$,
S.~Abdel~Khalek$^{\rm 115}$,
A.A.~Abdelalim$^{\rm 49}$,
O.~Abdinov$^{\rm 11}$,
R.~Aben$^{\rm 105}$,
B.~Abi$^{\rm 112}$,
M.~Abolins$^{\rm 88}$
% ...
\bigskip

% ...
$^{11}$ Institute of Physics, Azerbaijan Academy of Sciences, Baku, Azerbaijan\\
$^{12}$ Institut de F\'{i}sica d'Altes Energies and Departament de F\'{i}sica de la Universitat Aut\`{o}noma de Barcelona and ICREA, Barcelona, Spain\\
% ...
$^{21}$ Physikalisches Institut, University of Bonn, Bonn, Germany\\
% ...
$^{48}$ Fakult\"{a}t f\"{u}r Mathematik und Physik, Albert-Ludwigs-Universit\"{a}t, Freiburg, Germany\\
$^{49}$ Section de Physique, Universit\'{e} de Gen\`{e}ve, Geneva, Switzerland\\
% ...
$^{88}$ Department of Physics and Astronomy, Michigan State University, East Lansing MI, United States of America\\
% ...
$^{105}$ Nikhef National Institute for Subatomic Physics and University of Amsterdam, Amsterdam, Netherlands\\
% ...
$^{111}$ Homer L. Dodge Department of Physics and Astronomy, University of Oklahoma, Norman OK, United States of America\\
$^{112}$ Department of Physics, Oklahoma State University, Stillwater OK, United States of America\\
% ...
$^{115}$ LAL, Universit\'{e} Paris-Sud and CNRS/IN2P3, Orsay, France\\
% ...
\end{flushleft}
 
%
\end{document}